\documentclass[a4paper]{article} 
\usepackage{amsmath,amsthm,amssymb}

\title{Quantization of classical integrable systems \\
Part I: quasi-integrable quantum systems}
\author{M. Marino and N. N. Nekhoroshev \\
{\small Dipartimento di Matematica, Universit\`{a} degli Studi di Milano,} \\
{\small via Saldini 50, I-20133 Milano (Italy)}}

\newtheorem{thm}{Theorem}
\newtheorem{cor}[thm]{Corollary}
\newtheorem{lem}[thm]{Lemma}
\newtheorem{prop}[thm]{Proposition}

\theoremstyle{definition}
\newtheorem{defn}{Definition}[section]

\theoremstyle{remark}
\newtheorem{rem}{Remark}[section]

\def\beq{\begin{equation}}
\def\eeq{\end{equation}}

\def\ord{{\rm ord}\,}
\def\rank{{\rm rank}\,}

\def\dominio{K\times \mathbb{R}^n_p}

\def\mop{{\cal M}}
\def\aop{{\cal A}}
\def\bop{{\cal B}}

\def\hop{{\cal H}}
\def\sop{{\cal S}}
\def\wop{{\cal W}}
\def\yop{{\cal Y}}
\def\fop{{\cal F}}
\def\rn{\mathbb{R}}

\def\zn{\mathbb{Z}}

\def\gop{{\cal G}}
\def\oop{{\cal O}}

\def\zop{{\cal Z}}
\def\ipn2{\left[\frac n2\right]}

\numberwithin{equation}{section} \numberwithin{thm}{section}

\begin{document}

\maketitle

\begin{abstract}
We propose in this work a concept of integrability for quantum
systems, which corresponds to the concept of noncommutative
integrability for systems in classical mechanics. We determine a
condition for quantum operators which can be a suitable
replacement for the condition of functional independence for
functions on the classical phase space. This condition is based on
the properties of the main parts of the operators with respect to
the momenta. We are led in this way to the definition of what we
call a ``quasi-integrable quantum system''. This concept will be
further developed in a series of following papers.
\end{abstract}


\section{Introduction}

The theory of quantum systems is similar under many respects to
the theory of classical systems. Although there exist some
intrinsic differences, many important quantities, such as energy,
momentum, angular momentum, potential energy and so on, which play
a fundamental role in classical mechanics, are also usefully
employed in quantum mechanics after some modifications. To
quantities of this type, in the classical case there correspond
functions on the phase space of the system. The modification for
the quantum case is the following. The state of the system is no
longer described by a point in phase space, but by a function
defined on configuration space, called ``wave function''.
Furthermore, by means of some recipe, one associates with a
classical function on phase space a linear differential operator
acting on the wave function of the system. This correspondence
between classical functions and quantum operators is called
quantization. One takes as the analogue of classical Poisson
brackets between functions the commutators of the corresponding
linear operators \cite{Landau, Landau2}.

One of the main concepts of classical mechanics is that of
integrable system. There are several equivalent definitions. In
our context the following definition of hamiltonian integrable
system will be suitable \cite{TMMO}. Let us assume that the phase
space $M^{2n}$ is a $2n$-dimensional symplectic manifold, where
$n$ is called number of degrees of freedom. We say that the system
is integrable if there exists a set of functionally independent
real functions $F= (F_1,\ldots, F_k;F_{k+1},\ldots,F_{2n-k})$
defined on the phase space, with $1\leq k\leq n$, and this set has
the following properties. The first $k$ functions are in
involution with all functions of the set $F$, and $H$ is a
function of the $k$ ``central'' functions $F_1,\ldots,F_k$, i.e.
\begin{align}
\{F_i,F_j\}&=0\,,\qquad i=1,\ldots,k\,,\quad j=1,\ldots,2n-k \,,
\label{central} \\
H&=f(F_1,\ldots,F_k) \ . \label{ham}
\end{align}
Here $\{\,,\}$ denote Poisson brackets, and $f$ is an arbitrary
real function of $k$ variables. The most interesting case is the
compact case, i.e., the case in which some connected components of
the preimage $F^{-1}(b)$ of each point $b$ of the linear space
$\mathbb{R}^{2n-k}$ is a compact set. This case is investigated in
detail in \cite{TMMO}. The definition of integrability given above
is a generalization of the more usual one based on the well-known
Liouville--Arnold theorem \cite{Arnold}, which in the above
notation corresponds to the case $k=n$. For $k=n$ all functions of
the set $F$ are pairwise in involution, which is not true in the
cases with $k<n$. For this reason, the generalized concept of
integrability adopted here (see also \cite{fomenko, fasso}) is
sometimes called ``noncommutative integrability''. Some important
ideas, related with this type of integrability, can be extended in
a useful way also to significant classes of nonintegrable systems
\cite{Gordoc, Gord}.

The integrability in the classical case is based on two main
conditions: involution and functional independence. The latter
means that the differentials of the functions of the set $F$ are
linearly independent at almost all points of the phase space
$M^{2n}$. In section \ref{simplest} of the present paper some
well-known examples of classical integrable systems are reviewed,
together with their corresponding quantized systems.

The main aim of section \ref{firstp} is the construction of an
abstract definition of integrable quantum system, and the
determination of necessary conditions for integrability. In the
literature these issues have been approached in many different
ways (see for example \cite{berezin, cimapi, cipi, cigama,
semenov, enciso, hieta, cgm}). We shall adopt a definition of
integrable quantum system which is the direct analogue of the
definition of integrable classical system given above. Instead of
functions one considers linear differential operators of $n$
variables $x=(x_1, \ldots, x_n)$. To the product of functions
there corresponds the composition of operators. In the definition
of integrable system, the condition of involution is obviously
translated to the quantum case by changing the classical Poisson
brackets into commutators. However, the condition of functional
independence is apparently translated in a less trivial way
\cite{enciso, weigert, burch, chaly, hieta98, Gra_Wint}. We have
made an attempt to understand the meaning of this condition in
quantum mechanics. To this end, we have started from the concept
of algebraic dependence of a set of operators. In order to
formulate this condition in a way which is suitable for quantum
mechanics, we make use of the concept of ``main part'' of an
operator, i.e., its component of highest degree in the momenta.
Let us consider a quantum situation analogous to
(\ref{central})--(\ref{ham}). It means that a set of operators
${\cal F}= (\fop_1, \ldots, \fop_k; \fop_{k+1}, \dots,
\fop_{2n-k})$ satisfies the commutation relations $[\fop_i,
\fop_j]=0$ for $i=1,\ldots,k$, $j=1, \ldots, 2n-k$, and the
hamiltonian operator ${\cal H}$ of a quantum system is some
function of $\fop_1, \ldots, \fop_{k}$. We obviously assume that,
in case of algebraic dependence of the set ${\cal F}$, these
conditions do not ensure the integrability of the considered
quantum system. We are led in this way to introduce the notion of
``quasi-integrability'' of a quantum system. It is based on the
simple concept of ``quasi-independence'' of the set of operators
${\cal F}$, which is expressed as a property of the main parts of
these operators. We prove that the condition of quasi-independence
is sufficient to exclude algebraic dependence.

This paper is the first of a series of four, which are devoted to
integrable systems in classical and quantum mechanics. In the
second paper of this series (Part II) we will discuss about the
mathematical basis of the quantization by symmetrization. Then, in
Parts III and IV some classes of concrete integrable classical and
quantum systems will be described.

The problem of the correspondence between classical and quantum
mechanics has been discussed for a long time and appeared even
before the construction of modern quantum mechanics. The present
series of works represents an attempt to consider this
correspondence from a general perspective, with a particular
emphasis on the importance of the concept of noncommutatively
integrable system.

\section{Simplest examples of correspondence} \label{simplest}

Let us consider a few examples of integrable classical systems and
their corresponding quantum systems.

\subsection{The Kepler system} \label{kepler}

Let $(x,y,z)$ and $(r,\theta,\phi)$ respectively denote the
cartesian and polar coordinates in three-dimensional space. The
classical hamiltonian function for the Kepler system is
\begin{equation*}
H=\frac{\mathbf{p}^2}2 -\frac \alpha r =\frac{p_x^2 +p_y^2
+p_z^2}2 - \frac \alpha r = \frac 12 \left(p_r^2+ \frac
{p_\theta^2}{r^2} +\frac {p_\phi ^2} {r^2\sin^2\theta}\right)
-\frac \alpha r \ ,
\end{equation*}
where $\alpha$ is a real parameter. The configuration space $K$ is
the euclidean three-dimensional space without zero: $\mathbb {R}^3
\setminus \{{\bf 0}\}$, and the phase space $M=M^6$ is the
cotangent space $T^*K\sim \mathbb{R}^6\setminus \mathbb{R}^3$ to
$K$. The corresponding quantum hamiltonian operator on
$K=\mathbb{R}^3$ is
\begin{align*}
\hat H &=\frac{\mathbf{\hat p}^2}2 -\frac \alpha r =\frac{\hat
p_x^2 +\hat p_y^2 +\hat p_z^2}2 - \frac \alpha r \\
&= \frac 12 \left(\hat p_r^2- \frac{2i}{r}\hat p_r + \frac {1}
{r^2}\hat p_\theta^2 -i \frac{\cos \theta}{r^2 \sin\theta}\hat
p_\theta  +\frac {1} {r^2\sin^2\theta}\hat p_\phi^2 \right) -\frac
\alpha r \ ,
\end{align*}
where $\hat p_x=-i \hbar  \partial /\partial x$, $\hat p_r=-i
\hbar\partial /\partial r$, $\hat p_\theta=-i \hbar
\partial /\partial \theta$ and so on, with $i=\sqrt{-1}$. The classical
system has three integrals of motion which are the components of
the angular momentum vector $\mathbf M= \mathbf x \times \mathbf
p$. This implies that the orbit is planar. Another integral is for
example the angle $\psi$ between the longer axis of the elliptical
orbit in configuration space and a fixed axis in the plane of the
orbit. We thus obtain the integrable set of functions $(F_1; F_2,
F_3, F_4, F_5)= (H; M_1, M_2, M_3, \psi)$. For this system we then
have $n=3$ and $k=1$. For the quantum case the hamiltonian
operator $\hat H$ of the system commutes with the vector operator
of angular momentum $\mathbf {\hat M}= \mathbf {\hat x} \times
\mathbf {\hat p}$. Furthermore, there exists a quantum operator
which commutes with $\hat H$ and corresponds to the third
classical integral $\psi$. Therefore, also this integrable quantum
system has $k=1$. This is related to the fact that the set of
eigenvalues has one natural index, and all of these eigenvalues
(except the fundamental one) are degenerate.

\subsection{Rotations of a free rigid body} \label{rigid}

For this system one has $K=SO(3)$, $M^6=T^*SO(3)$. The hamiltonian
function $H$ of the system has the form
\[
H= \frac 12 \left(\frac {\Gamma_1^2}{I_1} + \frac
{\Gamma_2^2}{I_2}+ \frac {\Gamma_3^2}{I_3}\right)\,,
\]
where $I_1$, $I_2$, $I_3$ are the principal moments of inertia of
the rigid body, and $\Gamma_1$, $\Gamma_2$, $\Gamma_3$ are the
components of the angular momentum vector along the principal axes
of the body (see for example \cite{Landau, Arnold}). The system
has the integrals $M_x$, $M_y$, $M_z$, which are the components of
the angular momentum along the axes of a fixed inertial frame
$(x,y,z)$. If $SO(3)$ is parametrized by means of the usual Euler
angles $(\phi,\theta,\psi)$, we can write
\begin{align*}
\Gamma_1 &= \frac{\sin\psi}{\sin\theta} p_\phi +\cos\psi p_\theta
-\cot \theta\sin\psi p_\psi \,, \\
\Gamma_2 &= \frac{\cos\psi}{\sin\theta} p_\phi -\sin\psi p_\theta
-\cot \theta\cos\psi p_\psi \,, \\
\Gamma_3 &= p_\psi\,,
\end{align*}
and
\begin{align*}
M_x &= -\sin\phi\cot \theta p_\phi +\cos\phi p_\theta +
\frac{\sin\phi}{\sin\theta} p_\psi \,, \\
M_y &= \cos\phi\cot \theta p_\phi  +\sin\phi p_\theta-
\frac{\cos\phi}{\sin\theta} p_\psi \,, \\
M_z &= p_\phi\,.
\end{align*}
The system is integrable with $n=3$ and $k=2$, because almost
everywhere in $M^6$ a basis of the algebra of first integrals is
given by the set of functionally independent functions
$(F_1,F_2;F_3,F_4) =(H,\mathbf{M}^2;M_x,M_y)$. The corresponding
quantum system is obtained by taking as hamiltonian operator
\[
\hat H= \frac 12 \left(\frac {\hat \Gamma_1^2}{I_1} + \frac {\hat
\Gamma_2^2}{I_2}+ \frac {\hat \Gamma_3^2}{I_3}\right)\,,
\]
where
\begin{align*}
\hat\Gamma_1 &= \frac{\sin\psi}{\sin\theta} \hat p_\phi +\cos\psi
\hat p_\theta -\cot \theta\sin\psi \hat p_\psi \,, \\
\hat\Gamma_2 &= \frac{\cos\psi}{\sin\theta} \hat p_\phi -\sin\psi
\hat p_\theta-\cot \theta\cos\psi \hat p_\psi \,, \\
\hat\Gamma_3 &= \hat p_\psi\,,
\end{align*}
and $\hat p_\phi= -i\hbar\partial/\partial \phi$, etc.

\subsection{Symmetrical top in a gravitational field} \label{gravit}

The configuration and phase spaces for a rigid body in the
presence of a gravitational field are the same as for a free one.
However, the corresponding hamiltonian system is integrable only
if the rigid body has a symmetry axis. We have in this case
$n=k=3$, and a basis of the algebra of first integrals is given by
the set of functionally independent functions $(F_1,F_2,F_3) =(H,
M_z, \Gamma_3)$. Here $M_z$ is the component of the angular
momentum along the coordinate axis $z$ parallel to the direction
of the gravitational field, while $\Gamma_3$ is the component of
the angular momentum along the symmetry axis of the body. The
quantization is directly obtained as in the previous case.

\section{Quasi-integrable quantum systems} \label{firstp}

\subsection{Local and global independence of differential operators of finite order}
\label{loc}

Let $\hat p_i$ denote the differential operator $\hat p_i:=
\partial/\partial x_i$, $i=1, \ldots, n$. We denote the operator
of multiplication by a function $f(x)$ with the same symbol
$f(x)$. We consider functions $f \in C^\infty(K)$, where $K$ is an
open subset of $\mathbb{R}^n $. Typically $\mathbb{R}^n\setminus
K$ is a ``thin'' subset. For example, for the Kepler problem we
have $K=\mathbb {R}^3 \setminus \{{\bf 0}\}$, and for the two-body
problem $K=\mathbb {R}^6 \setminus \{{\mathbf x}={\mathbf y}\}$,
where ${\mathbf x}=(x_1,x_2,x_3)$, ${\mathbf y}= (y_1,y_2,y_3)$
are the position vectors of the two bodies in three-dimensional
space, and $({\mathbf x},{\mathbf y}) \in \mathbb {R}^6$.
%
%
Let us consider all the linear operators which are finite sums of
finite compositions of differential operators $p_i$, $i=1, \ldots,
n$, and arbitrary multiplication operators $f(x)$. The algebra of
such operators we denote by ${\cal O}= {\cal O}_K$. Note that we
are here considering only real operators.

Let us consider a linear differential operator ${\cal F}$ on
$K\subset\mathbb{R}^n $ of the form
\begin{equation}
{\cal F}=\sum_{|\alpha|\leq m}A_\alpha (x)\hat p^\alpha \ ,
\label{diff}
\end{equation}
where $m\in \mathbb{Z}_+$, $\alpha=(\alpha_1,\ldots,\alpha_n)\in
\mathbb{Z}_+^n$, $\mathbb{Z}_+ :=\{0,1,2,\ldots\}$, and
$|\alpha|:= \alpha_1+ \ldots +\alpha_n$. We have equivalently
$\mathbb{Z}_+^n := \mathbb{Z}^n\cap \mathbb{R}_+^n$,
$\mathbb{R}_+^n :=\{\alpha \in \mathbb{R}^n:\alpha_i \geq 0$ for
$i=1,\ldots,n\}$, where $\mathbb{Z}^n$ is the subsets of vectors
of $\mathbb{R}^n$ with integer components. Here $\hat
p^\alpha:=\partial^{|\alpha|}/
\partial x_1^{\alpha_1}\cdots \partial x_n^{\alpha_n}$. We
suppose that $A_\alpha(x)\in C^\infty (K)$ for any $\alpha$, that
is $A_\alpha(x)$ is an infinitely differentiable function on $K$,
$A_\alpha :K \to \mathbb{R}$.
\begin{defn} \label{identity}
The operators of the algebra ${\cal O}_K$ we call {\it operators
of class $\oop$}. We say that two operators ${\cal A}$ and ${\cal
B}$ of class $\oop$ are equal to each other if ${\cal A}\psi
={\cal B}\psi$ for any function $\psi \in C^\infty (K)$.
\end{defn}
\begin{prop}\label{rep}
An operator ${\cal F}$ of the form (\ref{diff}) is an operator of
class ${\cal O}$. Conversely, any operator of class ${\cal O}$ can
by represented in a unique way in the form (\ref{diff}).
\end{prop}
\begin{proof}
The first part of the proposition is obvious. The fact that an
operator of class ${\cal O}$ can be always represented in the form
(\ref{diff}) follows easily from the relation
\begin{equation} \label{leibniz}
\hat p_i f(x) = f(x)\hat p_i +\frac \partial{\partial x_i}f(x)\ ,
\end{equation}
which is a consequence of Leibniz rule of differentiation. In
order to prove that the representation of type (\ref{diff}) is
unique, let us consider the family of functions $\psi =e^{\lambda
x}$, $\lambda\in \mathbb{R}^n$, $\lambda x := \lambda_1 x_1+
\cdots + \lambda_n x_n$. It follows from (\ref{diff}) that
\begin{equation} \label{expo}
{\cal F}e^{\lambda x}= \sum_{|\alpha|\leq m}A_\alpha (x)
\lambda^\alpha e^{\lambda x} \ ,
\end{equation}
where $\lambda^\alpha:= \lambda_1^{\alpha_1} \cdots
\lambda_n^{\alpha_n}$. According to definition \ref{identity},
${\cal F}=0$ implies ${\cal F}e^{\lambda x}=0$ for every $\lambda
\in \mathbb{R}^n$ and every $x\in K$. Then from (\ref{expo}) we
derive that ${\cal F}=0$ if and only if $A_\alpha(x)=0$ for every
$\alpha \in \mathbb{Z}_+^n$ and every $x\in K$. This obviously
implies that a representation of the form (\ref{diff}) is unique
for any operator of class ${\cal O}$.
\end{proof}
\begin{defn} \label{standard}
We call formula (\ref{diff}) the {\it standard representation} of
the operator ${\cal F}$. The function $F:K\times
\mathbb{R}^{n}_{p} \to \mathbb{R}$ defined as
\begin{equation} \label{polyp}
F(x,p)=\sum_{|\alpha|\leq m}A_\alpha (x) p^\alpha\ ,
\end{equation}
where $p^\alpha := p_1^{\alpha_1}\cdots p_n^{\alpha_n}$, is called
the {\it symbol} of the operator ${\cal F}$. We shall also use the
notation ${\cal F}^{\rm smb}$ for $F$, that is ${\cal F}^{\rm
smb}=F$. It is a polynomial function of order $\leq m$ in the
variables $p=(p_1,\ldots, p_n)$.

Conversely, given a function $F:K\times \mathbb{R}^{n}_{p} \to
\mathbb{R}$ of the form (\ref{polyp}), we say that the
differential operator ${\cal F}$ on $K\subset\mathbb{R}^n $
defined by (\ref{diff}) is the {\it standard quantization} of $F$.
\end{defn}

\begin{defn}\label{hg}
Given an operator ${\cal F}$ of class ${\cal O}$ having the
standard representation (\ref{diff}), we call the {\it homogeneous
part of order $g\in \mathbb{Z}^+$ of the operator ${\cal F}$} the
operator
\[
H_g({\cal F}):= \begin{cases} \sum_{|\alpha|=g} A_\alpha(x)\hat
p^\alpha
&\mbox{if }0\leq g \leq m \\
0 \quad &\mbox{if }g>m \ .\end{cases}
\]
We define in a similar way the {\it homogeneous part $H_g(F)$ of
order $g$} of the symbol $F=F(x,p)$. Therefore, if $F={\cal
F}^{\rm smb}$ is the symbol of the operator ${\cal F}$, we have
$H_g(F)=(H_g({\cal F}))^{\rm smb}$.

If $\fop\neq 0$, let $m$ be maximum nonnegative integer such that
$H_m(\fop)\neq 0$. We call the operator
\[
M\fop :=H_m(\fop)
\]
the {\it main part} (with respect to $\hat p$) of the operator
${\cal F}$. Similarly, we call the function
\[
MF :=H_m(F)
\]
the {\it main part} (with respect to $p$) of the symbol $F$. We
call $m$ the {\it order} of the operator ${\cal F}$ and we write
$\ord {\cal F} =m$. If $\fop=0$, then we define $M\fop:=0$ and
$MF:=0$.
\end{defn}
Note that $MF$ is the symbol of the operator $M{\cal F}$, i.e.,
the main part of the symbol of an operator of class $\oop$
coincides with the symbol of its main part.

The following proposition can be easily proved by making use of
the identity (\ref{leibniz}).
\begin{prop} \label{prod}
If ${\cal A}=\sum_{|\alpha|\leq m_1}A_\alpha (x)\hat p^\alpha$,
${\cal B}=\sum_{|\alpha|\leq m_2}B_\alpha (x)\hat p^\alpha$ and
$a,b \in \mathbb R$, then
\begin{align*}
(a{\cal A} +b{\cal B})^{\rm smb} &=a A+b B \\
({\cal A}{\cal B})^{\rm smb} &= \sum_{0\leq|\alpha|\leq m_1} \frac
1 {\alpha !}\frac {\partial^{|\alpha|} A}{\partial p^\alpha}
\frac{\partial^{|\alpha|} B}{\partial x^\alpha}\ ,
\end{align*}
where $A={\cal A}^{\rm smb}$, $B={\cal B}^{\rm smb}$, $\alpha!
=\alpha_1! \cdots \alpha_n!$, $\partial x^\alpha =\partial
x_1^{\alpha_1} \cdots \partial x_n^{\alpha_n}$. We have for any
nonvanishing term of this sum
\[
\ord \frac 1 {\alpha !}\frac {\partial^{|\alpha|} A}{\partial
p^\alpha} \frac{\partial^{|\alpha|} B}{\partial x^\alpha} \leq m_1
+ m_2 -|\alpha|\ .
\]
\end{prop}

Let us consider a set of operators ${\cal F}_1,\ldots,{\cal F}_r$
of class ${\cal O}_K$. Let $MF_1,\ldots,MF_r$ be the main parts of
the symbols $F_1, \dots, F_r$ of these operators.

\begin{defn}[{\bf quasi-independence}] \label{strongind}
If the differentials $d(MF_1),\ldots,$ $d(MF_r)$ are linearly
independent at a point $(x,p) \in K\times \mathbb{R}^{n}_{p}$, we
will say that the operators ${\cal F}_1, \ldots,{\cal F}_r$ are
{\it quasi-independent} at that point. Moreover, if the
differentials $d(MF_1),\ldots,d(MF_r)$ are linearly independent at
almost each point of $K\times \mathbb{R}^{n}_{p}$, we will say
that the operators ${\cal F}_1, \ldots,{\cal F}_r$ are {\it
globally quasi-independent}.

For example, the operators $(x_1,\ldots,x_n, \hat p_1,\ldots, \hat
p_n)$ are globally quasi-independent. We call this set the {\it
standard set of operators}.

Similarly, if the differentials $d(MF_1),\ldots,d(MF_r)$ of the
main parts of $r$ functions $(F_1, \dots, F_r)$ are linearly
independent at a point $(x,p) \in K\times \mathbb{R}^{n}_{p}$, we
will say that these functions are {\it quasi-independent} at that
point. If the differentials $d(MF_1),\ldots,d(MF_r)$ are linearly
independent at almost each point of $K\times \mathbb{R}^{n}_{p}$,
we will say that these functions are {\it globally
quasi-independent}.
\end{defn}

The motivation of this definition of quasi-independence lies in
its connection with the property of dependence, which for a set of
operators will be formulated in section \ref{dep}. There we shall
show in fact that the property of quasi-independence defined above
is a sufficient condition to exclude dependence.

\subsection{Noncommutative polynomials}

In order to formulate a rigorous definition of ``dependence'' for
a set of operators, which in general do not commute with each
other, we need first of all to introduce the abstract notion of
``noncommutative polynomial''. In this section we will then define
the quasi-homogeneous parts and the main part of a noncommutative
polynomial with given weights and with respect to a given set of
operators. We will also give the definition of the homogeneous
parts of an operator.

Let us consider a set of real (commuting) variables $G=(G_1,
\ldots, G_l)$ and a set of ``noncommutative symbols'' $F=(F_1,
\ldots, F_r)$. At a later stage in this section we shall identify
the commuting variables $G$ with real multiplication operators,
and the noncommutative symbols $F$ with differential operators of
class $\oop$. However, we want here to establish formal properties
of noncommutative polynomials, which are independent of the
identification of the abstract variables $G$ and $F$ on which the
polynomials depend. Therefore, we shall here simply postulate that
formal operations of addition and multiplication are defined among
these variables, with the same properties of associativity and
distributivity which are valid for the corresponding operations
among operators. In particular, we shall postulate that
multiplication is commutative between any two variables $G$, and
noncommutative between any two variables $F$ or a variable $G$ and
a variable $F$. Moreover, a commutative multiplication is defined
between scalar numbers and variables $F$ or $G$. By making use of
these abstract operations, it is possible to define the formal
algebra of noncommutative polynomials in the following way.

\begin{defn}
If $\beta=(\beta_1, \ldots, \beta_q)$ is a finite sequence of
indexes, with $q\in {\mathbb N}$ and $\beta_i \in \{1, \ldots,
r\}$ for $i=1,\ldots, q$, we call {\it noncommutative monomial
(with respect to $F_1, \ldots, F_r$) associated with $\beta$} a
formal product
\begin{equation}
M_\beta =M_\beta(G,F) =Z_0 F_{\beta_1}Z_1 F_{\beta_2}\cdots
Z_{q-1} F_{\beta_q}Z_q \ , \label{ncmon}
\end{equation}
where $Z_i=Z_i(G)$ is an arbitrary usual function of class
$C^\infty$ of the commuting variables $G_1, \ldots, G_l$ for
$i=0,\ldots, q$. Note that some (or all) functions $Z_i$ can be
constants, for example $Z_i=1$. In particular, if $l=0$ then
monomials have the form $M_\beta(F)= cF_{\beta_1}F_{\beta_2}\cdots
F_{\beta_q}$, where $c$ is a constant. A function
$M_\emptyset=Z(G)$ independent of $F$ can be considered as a
noncommutative monomial associated with $\beta=\emptyset$.
\end{defn}

We consider that the monomial $M_\beta$ is (identically) {\it
zero} if and only if there exists $j \in \{0, \ldots, q\}$ such
that $Z_j$ is the zero function. The multiplication of a monomial
by a scalar and the product of two monomials are defined in an
obvious way. In particular, the product of two monomials
respectively associated with $\beta=(\beta_1, \ldots, \beta_q)$
and $\gamma=(\gamma_1, \ldots, \gamma_p)$ provides a monomial
associated with $(\beta_1, \ldots, \beta_q, \gamma_1, \ldots,
\gamma_p)$. This product is associative and noncommutative. We
further introduce the formal sum of noncommutative monomials,
which is assumed to be associative and commutative.

\begin{defn}\label{noncp}
We call {\it noncommutative polynomial} a formal sum of
noncommutative monomials. The set of such polynomials we denote
equivalently as ${\cal S}_N= {\cal S}_N^s= {\cal S}_N^{l,r} ={\cal
S}_N^{l,r}[G_1, \ldots, G_l, F_1, \ldots, F_r]$, where $s=l+r$. We
postulate that the operations of sum, multiplication by a scalar
and product among noncommutative polynomials enjoy all the usual
formal properties which make ${\cal S}_N^s$ a noncommutative
algebra. We call {\it noncommutative polynomial associated with
$\beta$} a formal sum of monomials all associated with the same
$\beta$, that is an expression
\begin{equation}
S_\beta= S_\beta(G,F) =\sum_{i\in I}M_{\beta,i} \ , \label{ncpolb}
\end{equation}
where $I$ is a finite set of indexes and $M_{\beta,i}$ is a
noncommutative monomial associated with $\beta$ for all $i\in I$.
\end{defn}

We consider that the polynomial (\ref{ncpolb}) is (identically)
{\it zero} if the equality $\sum_i M_{\beta,i}= 0$ follows from
the formal properties of the algebraic operations. Obviously, two
polynomials are considered to be equal to each other if and only
if their difference is zero. For example, we have $Z_1 F_1 Z_2 F_2
+ Y_1 F_1 Y_2 F_2 = X_1 F_1 Z_2 F_2 + Y_1 F_1 X_2 F_2$ if $X_1(G)=
Z_1(G) -Y_1(G)$ and $X_2(G)= Z_2(G)+ Y_2(G)\,\forall\, G\in
\rn^l$.

For $Q\in {\mathbb Z}_+$, let $B_Q$ denote the set of all finite
sequences $\beta= (\beta_1, \ldots, \beta_q)$, with $0\leq q\leq
Q$ and $\beta_i \in \{1, \ldots, r\}$ for $i=1,\ldots, q$. A
generic noncommutative polynomial can then be expressed in the
form
\begin{equation}
S=\sum_{\beta\in B_Q}S_\beta \ , \label{ncpol}
\end{equation}
where $S_\beta$ is a noncommutative polynomial associated with
$\beta$ for each $\beta\in B_Q$. We consider that $S$ is
(identically) zero if and only if $S_\beta \equiv 0$ for all
$\beta\in B_Q$.

\begin{defn}\label{abel}
With any noncommutative polynomial $S(G,F)$ we associate the
commutative polynomial $S_C(G,F)$ which is obtained by considering
all variables $G_1, \ldots,G_l$, $F_1,\ldots, F_r$ as commuting
variables, and then operating the reduction of analogous terms,
i.e., monomials of $F_1, \ldots, F_r$ which differ from each other
only at most in the coefficients. (These coefficients are
functions of $G_1, \ldots, G_l$, or constants if $l=0$.) We call
the transformation
\begin{equation*}
T:S\mapsto S_C
\end{equation*}
the {\it abelianization} of $S$.
\end{defn}

\begin{defn}\label{hom}
Let $w=(w_1, \ldots, w_r)$ be a given set of natural numbers. If
the noncommutative polynomial $S\in \sop^{l,r}_N$ is represented
in the form (\ref{ncpol}), let us consider for $d\in {\mathbb
Z}_+$ the subset of $B_Q$
\[
{\cal L}_{d,w} := \bigg\{\beta= (\beta_1, \ldots, \beta_q) \in
B_Q:\ \sum_{j=1}^{q}w_{\beta_j} =d \bigg\} \ .
\]
We will call the polynomial
\begin{equation*}
C_d=C_{d,w}(S) :=\sum_{\beta \in {\cal L}_{d,w}}S_\beta
\end{equation*}
the {\it quasi-homogeneous part of degree $d$ with weights $w$ of
the polynomial $S$}. If $S=C_{d,w}(S)\neq 0$, then $S$ is called a
{\it quasi-homogeneous polynomial of degree $d$ with weights $w$}.
If $S\neq 0$, we define the {\it degree $\deg_w S$ of the
polynomial (\ref{ncpol}) with weights $w$} the maximum $d\in
\zn_+$ such that $C_{d,w}(S)\neq 0$. We define the {\it main part
${\cal M}_{\cal F}S$ of the polynomial $S$ with weights $w$} as
\begin{equation*}
{\cal M}_{w}S:=C_{\bar d,w}(S)\ ,
\end{equation*}
where $\bar d=\deg_w S$. If $S=0$, then we define $\mop_w S:=0$.

We define the {\it main part ${\cal M}_{\cal F}S$ of the
polynomial $S$ with respect to the set ${\cal F}=({\cal F}_1,
\ldots, {\cal F}_r)$} of operators of class ${\cal O}$ as
\[
{\cal M}_{\cal F}S={\cal M}_{w}S\,, \quad
\mbox{where }w=(w_1, \ldots, w_s)\,,\quad w_i=\ord {\cal F}_i
\mbox{ for } i=1,\ldots,r\,.
\]
We will also write $\deg_{\cal F}S:=\deg_w S= \bar d$, and we say
that $\bar d$ is the {\it degree of $S$ with respect to ${\cal
F}$}. If ${\cal W}=({\cal G},{\cal F})$, where $\gop= (\gop_1,
\ldots,\gop_l)$ is a set of multiplication operators (i.e., a set
of operators of class $\oop$ such that $\ord \gop_i=0\, \forall\,
i=1, \dots, l$), we shall also write ${\cal M}_{\cal W}S={\cal
M}_{\cal F}S$.
\end{defn}

\begin{lem} \label{prop3} Let $S$ be a noncommutative polynomial.
If $S_C\neq 0$ then $S\neq 0$. If $S$ is a quasi-homogeneous
polynomial of degree $d$ with weights $w$, then either $S_C=0$ or
$S_C$ is also a quasi-homogeneous polynomial of degree $d$ with
weights $w$.
\end{lem}
Using this obvious lemma and proposition \ref{prod} it is easy to
prove the following proposition.
\begin{prop} \label{mainp}
Let ${\cal W}=({\cal G},{\cal F})$ be a set of operators of class
${\cal O}$, where ${\cal G}= ({\cal G}_{1}, \ldots, {\cal
G}_{l})$, $\ord \gop_i=0\, \forall\, i=1, \dots, l$, and ${\cal F}
=({\cal F}_1,\ldots,{\cal F}_{r})$. Let $G$ be the symbol of
$\gop$, and $MF$ the symbol of the main parts of $\fop$. Let $S\in
{\cal S}_N^{l,r}$ be a noncommutative polynomial such that
$\deg_{{\cal W}}S =\bar d$. Then $S({\cal G},{\cal F}):=
S(G,F)|_{G={\cal G}, F={\cal F}}$ is an operator of class ${\cal
O}$, and
\begin{equation}\label{top}
\big(H_{\bar d}(S({\cal G,F}))\big)^{\rm smb}= ({\cal M}_{{\cal
W}}S)_C (G,MF) \ .
\end{equation}
\end{prop}
The above formula means that, if the noncommutative polynomial $S$
has degree $\bar d$ with respect to $\wop$, then the symbol of the
homogeneous part of order $\bar d$ of $S({\cal G,F})$ is obtained
by taking the symbols $(G,MF)$ as the variables of the
abelianization of the main part of $S$ with respect to $\wop$.
Note that the two members of equality (\ref{top}) may be zero.
When they are nonzero, they represent the symbol of the main part
of the operator $S({\cal G},{\cal F})$.

\subsection{Dependent sets of operators} \label{dep}

In this section we define the concept of dependence for a set of
operators. There seems to exist no definition as general and
natural as that for the functional dependence of a set of
functions. Our idea is here to consider as dependent all sets of
operators for which there exists a correlation, that is a
nontrivial operator function which, when its arguments are
replaced by the operators of the set, vanishes identically in a
neighborhood of a given point of configuration space. However,
when we are dealing with operators, the class of available
functions in our general scheme is restricted to noncommutative
polynomials. Hence such a definition of dependence would appear
too restrictive. We shall therefore adopt a different definition,
which offers a higher level of generality (see definition
\ref{implic}). We will prove that a dependent set defined in this
way cannot be quasi-independent (see theorem \ref{indep}).

We begin by giving the definition of ``regular correlation'' among
the class ${\cal O}$ operators of the set ${\cal W} =({\cal G}_1,
\ldots, {\cal G}_l,$ ${\cal F}_1, \ldots, {\cal F}_r)$, with $\ord
{\cal G}_i =0$ for $i=1,\ldots, l$, and $\ord {\cal F}_j \geq 1$
for $j=1,\ldots, r$. Let $MW: K\times \mathbb{R}_p^n \to
\mathbb{R}^{l+r}$ denote the vector function $(G_1, \ldots, G_l,
MF_1, \ldots, MF_r)$ defined by the symbols of the main parts of
these operators. Let us consider a point $(\bar x, \bar p)\in
K\times \mathbb{R}_p^n$ and its image $\bar W :=MW(\bar x, \bar p)
\in \mathbb{R}^{l+r}$ with respect to the function $MW$. Let us
suppose that there exists a nonvanishing noncommutative polynomial
$S=S(G,F) \in {\cal S}_N^{l,r}$ with the following two properties:
\begin{enumerate}

\item The differential of the abelianization $({\cal M}_{\cal W}
S)_C$ of the main part ${\cal M}_{\cal W}S$ of the noncommutative
polynomial $S$ with respect to ${\cal W}$ is nonzero at $\bar W$,
that is
\begin{equation}
d({\cal M}_{\cal W} S)_C(\bar W)\neq 0 \ . \label{comm}
\end{equation}

\item $S({\cal G},{\cal F}):= S(G,F)|_{G={\cal G}, F={\cal F}}$ is
the zero operator in a neighborhood $H\subset K$ of $\bar x$, that
is
\begin{equation}
S({\cal G},{\cal F})\psi(x)= 0
\end{equation}
$\forall\,\psi\in C^\infty (H)$ and $\forall \,x\in H$.
\end{enumerate}
Note that, according to propositions (\ref{rep}) and
(\ref{mainp}), property 2 implies that $({\cal M}_{\cal W}
S)_C(\bar W) =0$.

\begin{defn} \label{correl}
We say that a noncommutative polynomial $S\in {\cal S}_N^{l,r}$
with the properties 1 and 2 is a {\it regular correlation among
the operators ${\cal W}$ at the point $(\bar x, \bar p)$}. We say
that the correlation is {\it global} if it is a correlation at
almost all points of $K\times \mathbb{R}_p^n$.
\end{defn}

In order to understand the meaning of condition (\ref{comm}) in
the above definition, consider the space $K=\mathbb{R}$ and the
two operators ${\cal G}=x$, ${\cal F}=\hat p$. If condition
(\ref{comm}) were not required, the noncommutative polynomial
$S(G,F)= GF-FG +1$ would be a global regular correlation between
the operators ${\cal G}$ and ${\cal F}$, but this would be in
contradiction with the fact that these operators are
quasi-independent according to definition \ref{strongind}.

As we have already explained, our aim is to find a definition of
dependence which includes (but is not restricted to) the case in
which there exists a regular correlation among the operators of a
set. As a first step in this direction, we now introduce the
concept of dependence of a set of operators on another set. To
this purpose, let us consider two sets of operators ${\cal W}=
({\cal W}_1, \ldots, {\cal W}_s)$ and ${\cal Y}= ({\cal Y}_1,
\ldots, {\cal Y}_r)$ of class ${\cal O}$. We can suppose in
general that $\ord \wop_i=0$ and $\ord \yop_j=0$ for $i=1, \dots,
s'$ and $j=1, \dots, r'$, where $0\leq s' \leq s$ and $0\leq r'
\leq r$. Let $(MW,MY): K\times \mathbb{R}_p^n \to
\mathbb{R}^{s+r}$ be the vector function defined by the symbols of
the main parts of these operators, and $(\bar W,\bar Y):=
(MW,MY)(\bar x, \bar p) \in \mathbb{R}^{s+r}$ the image of the
point $(\bar x, \bar p)$ with respect to this function. Let us
consider the class of noncommutative polynomials ${\cal
S}_N^{s+r}= {\cal S}_N^{l,m}$, where $l=s'+r'$ and $m=s+r -l$. Let
us suppose that there exists a set of noncommutative polynomials
$S=(S_1, \ldots, S_s)$ of this class with the following two
properties:
\begin{enumerate}

\item Let ${\cal M}_{({\cal W},{\cal Y})} S$ be the set of main
parts ${\cal M}_{({\cal W},{\cal Y})} S_i$ of the components $S_i$
of the vector $S$, with $i=1,\ldots, s$, and let $({\cal
M}_{({\cal W},{\cal Y})} S)_C$ be the vector of their
abelianizations. Then
\[
\det\frac{\partial ({\cal M}_{({\cal W},{\cal Y})} S)_C}{\partial
W} (\bar W,\bar Y)\neq 0\ ,
\]
where on the left-hand side we have the determinant of the
$s\times s$ Jacobi matrix with respect to $W$ of the vector
function $({\cal M}_{({\cal W},{\cal Y})} S)_C$ of the $s+r$
variables $(W,Y)$.

\item  After substitution of the $s+r$ operators $({\cal W},{\cal
Y})$ to the variables of the polynomial $S$ we obtain the zero
operator in a neighborhood $H\subset K$ of $\bar x$: $S({\cal
W},{\cal Y})=0$ in $H$.
\end{enumerate}
Owing to propositions (\ref{rep}) and (\ref{mainp}), property 2
clearly implies that

\noindent$({\cal M}_{({\cal W},{\cal Y})} S)_C(\bar W, \bar Y)
=0$. Note also that the $S_i$ are correlations at $(\bar x, \bar
p)$ among the operators $({\cal W},{\cal Y})$ in the sense of
definition \ref{correl}.
\begin{defn} \label{contain}
In this case we say that the set of operators ${\cal W}$ {\it is
algebraically dependent} on the set of operators ${\cal Y}$ at the
point $(\bar x, \bar p)$.
\end{defn}
\begin{rem} \label{selfcont0}
Obviously, if the set of operators ${\cal W}$ is (properly or
improperly) contained in the set ${\cal Y}$, i.e., ${\cal W}
\subseteq {\cal Y}$, then ${\cal W}$ is algebraically dependent on
${\cal Y}$ at any point $(x,p)\in K\times \mathbb{R}_p^n$. In
fact, suppose that $s\leq r$ and ${\cal W}_i={\cal Y}_i\ \forall
\, i=1, \dots, s$. Then the vector of polynomials $S_i(W,Y)= W_i
-Y_i$ for $i=1,\ldots,s$ clearly satisfies the conditions 1 and 2
of definition \ref{contain}. It follows that, in particular, any
set of operators ${\cal W}$ is algebraically dependent on itself
at any point $(x,p)\in K\times \mathbb{R}_p^n$. It is also
immediate to see that, if ${\cal W}= ({\cal W}_1, \ldots, {\cal
W}_s)$ and ${\cal W}'= ({\cal W}_1, \ldots, {\cal W}_s, {\cal
W}_{s+1})$, where ${\cal W}_{s+1} ={\cal W}_j$ for some $j$ with
$1\leq j\leq s$, then ${\cal W}'$ is algebraically dependent on
${\cal W}$ and ${\cal W}$ is algebraically dependent on ${\cal
W}'$.
\end{rem}

A particularly simple type of algebraic dependence occurs when
(some powers of) the operators of the set $\wop$ can be explicitly
expressed as noncommutative polynomial functions of the operators
of the set $\yop$. This is stated in more precise terms by the
following proposition.
\begin{prop} \label{example}
Let us consider the set of operators ${\cal W} =({\cal W}_1,
\ldots, {\cal W}_s)$. Let us suppose that there exists another set
of operators ${\cal Y} =({\cal Y}_1, \ldots, {\cal Y}_r)$ such
that
\begin{equation}
{\cal W}_i^{j_i}=S_i({\cal Y})\ , \label{corr}
\end{equation}
where $j_i\in {\mathbb N}$ (for example $j_i=1$) and $S_i({\cal
Y})$ is a noncommutative polynomial of class ${\cal S}_N^r$ such
that
\begin{equation} \label{cond3}
\deg_{\cal Y}S_i=j_i\,\ord {\cal W}_i\quad \forall \,i=1,\ldots,
s\,.
\end{equation}
Then ${\cal W}$ is algebraically dependent on ${\cal Y}$ at all
points $(x,p)$ such that $MW_i(x,p) \neq 0\, \forall\, i\in J$,
where $J:=\{i\in \{1,\ldots, s\}: j_i>1\}$. In particular, ${\cal
W}$ is algebraically dependent on ${\cal Y}$ in all $\dominio$ if
$j_i=1\, \forall\, i$.
\end{prop}
\begin{proof}
Let $\bar{S}_i(W,Y):= W_i^{j_i}-S_i(Y)$ denote the given
correlations among the operators ${\cal W}$ and ${\cal Y}$. The
element with pair of indexes $(i,h)$ of the Jacobi matrix for the
abelianization of the main part of $\bar{S}$ is
\begin{equation}\label{ex1}
\frac\partial{\partial W_h}({\cal M}_{({\cal W,Y})} \bar{S}_i)_C =
\delta_{ih}j_i W_i^{j_i -1}\,.
\end{equation}
Hence this Jacobi matrix is diagonal (in the previous formula
$\delta_{ih}$ is the Kr\"onecker symbol, such that
$\delta_{ii}=1$, $\delta_{ih}=0$ for $i\neq h$) and nondegenerate
for $W=MW(x,p)$ if $MW_i(x,p) \neq 0\, \forall\, i\in J$.
Therefore at such points $(x,p)$ both conditions of definition
\ref{contain} are satisfied.

Note that condition (\ref{cond3}) is necessary for the validity of
formula (\ref{ex1}). In fact, if this condition is not true for
some $i$, then clearly $\ord{\cal W}_i^{j_i} <\deg_{\cal Y}
S_i({\cal Y})$ for this $i$, so that ${\cal M}_{({\cal W,Y})}
\bar{S}_i= {\cal M}_{({\cal Y})} S_i$. It follows that all the
elements of the $i$-th row of the Jacobi matrix for $({\cal
M}_{({\cal W,Y})} \bar{S})_C$ are identically zero. Hence the
determinant of this matrix is also identically zero.
\end{proof}

\begin{prop} \label{wxp}
Let ${\cal W} =({\cal W}_1, \ldots, {\cal W}_s)$ be a set of
operators of class ${\cal O}_K$. Then $\wop$ is algebraically
dependent on $(x,\hat p)$ in all $\dominio$, where $(x, \hat p)
=(x_1, \dots, x_n, \hat p_1, \dots, \hat p_n)$ is the standard set
of operators.
\end{prop}
\begin{proof}
Let
\[
{\cal W}_j= \sum_{|\alpha|\leq m_j}A_{j,\alpha} (x)\hat p^\alpha \
,\quad j=1,\ldots, s \,,
\]
be the standard representation of the operators ${\cal W}_j\in
{\cal O}$, see definition \ref{standard}. We can rewrite these
equalities as
\[
{\cal W}_j= S_j(x, \hat p)\,,\quad j=1,\ldots,s \,,
\]
where $S_j:=\sum_{|\alpha|\leq m_j}A_{j,\alpha} (X)P^\alpha$ is a
noncommutative polynomial of class $\sop_N^{n,n}$ for $j= 1,
\ldots, s$. Clearly
\[
\ord {\cal W}_j= \deg_{(x,\hat{p})}S_j \quad \forall \,j=1,\ldots,
s \,.
\]
The thesis then follows from proposition \ref{example}.
\end{proof}

\begin{prop} \label{union0}
If ${\cal W}^{(1)}$ is algebraically dependent on ${\cal Y}^{(1)}$
and ${\cal W}^{(2)}$ is algebraically dependent on ${\cal
Y}^{(2)}$ at $(\bar x, \bar p)$, then at that point ${\cal W}$ is
algebraically dependent on ${\cal Y}$, where ${\cal W} := {\cal
W}^{(1)} \cup {\cal W}^{(2)}$ and ${\cal Y}:= {\cal Y}^{(1)} \cup
{\cal Y}^{(2)}$.
\end{prop}
\begin{proof}
If ${\cal W}^{(i)}= ({\cal W}_1^{(i)}, \ldots, {\cal
W}_{s_i}^{(i)})$ and ${\cal Y}^{(i)}= ({\cal Y}_1^{(i)}, \ldots,
{\cal Y}_{r_i}^{(i)})$, for $i=1,2$, according to the hypotheses
there exist two sets of noncommutative polynomials $S^{(i)}=
(S_1^{(i)}, \ldots, S_{s_i}^{(i)})$ of class ${\cal
S}_N^{s_i+r_i}$ such that
\[
\det\frac{\partial ({\cal M}_{({\cal W}^{(i)}, {\cal Y}^{(i)})}
S^{(i)})_C}{\partial W^{(i)}} (\bar W^{(i)},\bar Y^{(i)}) \neq 0\
\]
and
\[
S^{(i)}({\cal W}^{(i)},{\cal Y}^{(i)})=0 \quad \mbox{in } H_i \ ,
\]
where $(\bar W^{(i)},\bar Y^{(i)}):= (MW^{(i)},MY^{(i)})(\bar x,
\bar p) \in \mathbb{R}^{s_i+r_i}$ and $H_i\subset K$ is an open
neighborhood of $\bar x$. Then for the set
\[
S= (S_1^{(1)}, \ldots, S_{s_1}^{(1)},S_1^{(2)}, \ldots,
S_{s_2}^{(2)})
\]
of $s$ noncommutative polynomials of class ${\cal S}_N^{s+r}$,
where $s=s_1 +s_2$ and $r=r_1 +r_2$, we have
\begin{align*}
\det\frac{\partial ({\cal M}_{({\cal W},{\cal Y})} S)_C}{\partial
W} (\bar W,\bar Y) = &\det\frac{\partial ({\cal M}_{({\cal
W}^{(1)}, {\cal Y}^{(1)})} S^{(1)})_C}{\partial W^{(1)}} (\bar
W^{(1)},\bar Y^{(1)}) \\
&\times\det\frac{\partial ({\cal M}_{({\cal W}^{(2)}, {\cal
Y}^{(2)})} S^{(2)})_C}{\partial W^{(2)}} (\bar W^{(2)},\bar
Y^{(2)}) \neq 0
\end{align*}
and
\[
S({\cal W},{\cal Y})=0 \quad \mbox{in } H= H_1 \cap H_2 \ ,
\]
where
\begin{align*}
{\cal W}&= ({\cal W}_1^{(1)}, \ldots, {\cal W}_{s_1}^{(1)}, {\cal
W}_1^{(2)}, \ldots, {\cal W}_{s_2}^{(2)}) \\
{\cal Y}&= ({\cal Y}_1^{(1)}, \ldots, {\cal Y}_{r_1}^{(1)}, {\cal
Y}_1^{(2)}, \ldots, {\cal Y}_{r_2}^{(2)}) \\
(\bar W,\bar Y)&= (MW,MY)(\bar x, \bar p) \in \mathbb{R}^{s+r}\ .
\end{align*}
This means that ${\cal W}$ is algebraically dependent on ${\cal
Y}$.
\end{proof}

The following important proposition says that, if the set $\wop$
is algebraically dependent on the set $\yop$ at a point of
$\dominio$, then the same relation of algebraic dependence also
holds in a full neighborhood of this point. Furthermore, in this
neighborhood the symbols $MW$ of the main parts of $\wop$ are
functionally dependent on the symbols $MY$ of the main parts of
$\yop$.
\begin{prop}\label{chiave}
Let ${\cal W}= ({\cal W}_{1}, \ldots, {\cal W}_{s})$ and ${\cal
Y}=({\cal Y}_1,\ldots,{\cal Y}_{r})$ be two sets of operators of
class ${\cal O}_K$, such that ${\cal W}$ is algebraically
dependent on ${\cal Y}$ at $(\bar x, \bar p)\in \dominio$. Then
there exists a neighborhood $O\subset \dominio$ of $(\bar x, \bar
p)$ such that:
\renewcommand{\theenumi}{\roman{enumi}}
\begin{enumerate}
\item The main parts $MW$ of the symbols of $\wop$ are functions
of the main parts $MY$ of the symbols of $\yop$ in $O$. More
precisely, there exist a neighborhood $\Omega\subset \rn^r$ of
$\bar Y= MY(\bar x, \bar p)$, and a function $f:\Omega \to \rn^s$,
$f\in C^\infty(\Omega)$, such that $MW(x,p)= f(MY(x,p)) \ \forall
\,(x,p)\in O$.

\item ${\cal W}$ is algebraically dependent on ${\cal Y}$ at all
points of $O$.
\end{enumerate}
\renewcommand{\theenumi}{\arabic{enumi}}
\end{prop}
\begin{proof}
Let $S=(S_1, \ldots, S_s)$ be the vector of noncommutative
polynomials of class ${\cal S}_N^{s+r}$ satisfying conditions 1
and 2 of definition \ref{contain}. Since, by condition 2,
$S_i({\cal W,Y})$ is the zero operator in a neighborhood $H\subset
K$ of $\bar x$ for all $i=1, \dots, s$, its homogeneous part
$H_{\bar d_i}(S_i({\cal W,Y}))$ of order $\bar d_i$ is also
obviously zero in $H$, where $\bar d_i= \deg_{(\wop, \yop)}S_i$.
Therefore, using proposition \ref{mainp} we obtain
\begin{equation} \label{ch1}
\tilde S((MW,MY)(x,p)) =0 \quad \forall\, (x,p) \in
H\times\mathbb{R}^n_p\,,
\end{equation}
where we have introduced the vector function $\tilde S := ({\cal
M}_{({\cal W,Y})}S)_C$, $\tilde S:\mathbb{R}^{s+r} \to \rn^s$.
This implies in particular that $\tilde S (\bar W,\bar Y) =0$.
According to our definition of noncommutative polynomial, we know
that $\tilde S\in C^\infty (\mathbb{R}^{r+s})$. Taking also into
account condition 1 of definition \ref{contain}, it follows from
the theorem on implicit functions that there exist a neighborhood
$\Omega\subset \rn^r$ of $\bar Y$, a neighborhood $\Omega'\subset
\rn^s$ of $\bar W$, and a function $f:\Omega \to \Omega'$, $f\in
C^\infty (\Omega)$, with the following properties:
\renewcommand{\theenumi}{\alph{enumi}}
\begin{enumerate}
\item $f(\bar Y)= \bar W$.

\item $\tilde S (f(Y),Y)=0\ \forall\, Y\in\Omega$.

\item For all $Y\in\Omega$, $W=f(Y)$ is the only solution $W\in
\Omega'$ of the equation $\tilde S (W,Y)=0$. \label{unica}

\item If $\partial \tilde S/\partial W$ and $\partial \tilde
S/\partial Y$ denote the Jacobi matrices of the vector function
$\tilde S$ with respect to variables $W$ and $Y$ respectively,
then
\begin{equation}\label{ch2}
\det\frac{\partial \tilde S}{\partial W}(W,Y) \neq 0\quad
\forall\, (W,Y)\in \Omega\times \Omega'
\end{equation}
and
\begin{equation}
df(Y)=-\left(\frac{\partial \tilde S}{\partial
W}(f(Y),Y)\right)^{-1} \left(\frac{\partial \tilde S}{\partial
Y}(f(Y),Y)\right) \quad \forall\, Y\in \Omega\,.
\end{equation}
\end{enumerate}
\renewcommand{\theenumi}{\arabic{enumi}}
According to our definition of operator of class ${\cal O}_K$, we
know that $(MW,MY)\in C^\infty(\dominio)$. Therefore there exists
a neighborhood $\tilde O\subset \dominio$ of $(\bar x, \bar p)$
such that
\begin{equation}\label{ch3}
(MW,MY)(x,p)\in \Omega\times \Omega' \quad \forall\, (x,p)\in
\tilde O \,.
\end{equation}
The set $O:= \tilde O\cap (H\times \rn^n_p)$ is also a
neighborhood of $(\bar x, \bar p)$.
From (\ref{ch1}) and from property \ref{unica} of function $f$, it
then follows that $MW(x,p)= f(MY(x,p)) \ \forall \,(x,p)\in O$.
Furthermore, from (\ref{ch2}) and (\ref{ch3}) it follows that
\begin{equation*}
\det\frac{\partial \tilde S}{\partial W}((MW,MY)(x,p)) \neq 0\quad
\forall\, (x,p)\in O\,.
\end{equation*}
Therefore, according to definition \ref{contain}, ${\cal W}$ is
algebraically dependent on ${\cal Y}$ at all points of $O$.
\end{proof}

The relation of dependence between sets of operators, given by
definition \ref{contain}, does not automatically enjoy the
transitive property. For this reason, we will now introduce
another relation which extends that of algebraic dependence in
such a way to insure transitivity.
\begin{defn} \label{contain2}
Let us suppose that there exist $q$ finite sets ${\cal Z}^{(1)},
\ldots, {\cal Z}^{(q)}$ of operators of class ${\cal O}$, such
that ${\cal W}$ is algebraically dependent on ${\cal Z}^{(1)}$ at
the point $(\bar x, \bar p)$, ${\cal Z}^{(i)}$ is algebraically
dependent on ${\cal Z}^{(i+1)}$ at $(\bar x, \bar p)$ for $i=1,
\ldots, q-1$, and ${\cal Z}^{(q)}$ is algebraically dependent at
$(\bar x, \bar p)$ on ${\cal Y}$. In this case we say that the set
of operators ${\cal Y}$ {\it algebraically contains} the set of
operators ${\cal W}$ at the point $(\bar x, \bar p)$ and we write
${\cal Y}\sqsupseteq{\cal W}$. We also say that the set ${\cal W}$
{\it is algebraically contained} in the set ${\cal Y}$ at $(\bar
x, \bar p)$ and we write ${\cal W}\sqsubseteq{\cal Y}$.
\end{defn}
\begin{rem} \label{selfcont}
Obviously, if ${\cal W}$ is algebraically dependent on ${\cal Y}$
at a point $(x,p)$, then ${\cal W}\sqsubseteq{\cal Y}$ at that
point. In particular, ${\cal W}\sqsubseteq{\cal W}$ at any point
$(x,p)\in K\times \mathbb{R}_p^n$ for any set of operators ${\cal
W}$. Furthermore, if ${\cal W}\sqsubseteq{\cal Y}$ and ${\cal
Y}\sqsubseteq{\cal Z}$ at a point $(x,p)$, then ${\cal
W}\sqsubseteq{\cal Z}$ at that point.
\end{rem}
The following proposition can be deduced from proposition
\ref{union0}.
\begin{prop} \label{union}
If ${\cal W}^{(1)}\sqsubseteq {\cal Y}^{(1)}$ and ${\cal
W}^{(2)}\sqsubseteq {\cal Y}^{(2)}$ at $(\bar x, \bar p)$, then at
that point ${\cal W}\sqsubseteq {\cal Y}$, where ${\cal W} :=
{\cal W}^{(1)} \cup {\cal W}^{(2)}$ and ${\cal Y}:= {\cal Y}^{(1)}
\cup {\cal Y}^{(2)}$.
\end{prop}
\begin{cor} \label{unionb}
If ${\cal W}^{(1)}\sqsubseteq {\cal Y}$ and ${\cal
W}^{(2)}\sqsubseteq {\cal Y}$ at $(\bar x, \bar p)$, then at that
point ${\cal W}\sqsubseteq {\cal Y}$, where ${\cal W} := {\cal
W}^{(1)} \cup {\cal W}^{(2)}$.
\end{cor}

\begin{prop}
Let ${\cal W}=({\cal W}_1, \ldots, {\cal W}_s)$, ${\cal V}=({\cal
V}_1, \ldots, {\cal V}_v)$ and ${\cal Y}=({\cal Y}_1, \ldots,
{\cal Y}_r)$ be three sets of operators of class $\oop$. Let us
consider a point $(\bar x, \bar p)\in K\times \mathbb{R}_p^n$, and
its image $(\bar W, \bar V, \bar Y)=(MW,MV,MY)(\bar x, \bar p)\in
\mathbb{R}^{s+v+r}$ with respect to the symbols of the main parts
of ${\cal W}$, ${\cal V}$, ${\cal Y}$. Let us suppose that there
exists a set $S=(S_1, \ldots, S_{s+v})$ of $s+v$ noncommutative
polynomials of class ${\cal S}_N^{s +v +r}$, such that
\[
\det \frac\partial {\partial(W,V)}({\cal M}_{({\cal W,V,Y})}S)_C
(\bar W, \bar V, \bar Y) \neq 0 \ ,
\]
and $S({\cal W},{\cal V},{\cal Y})=0$ in a neighborhood $H\subset
K$ of $(\bar x, \bar p)$, that is
\[
S({\cal W},{\cal V},{\cal Y})\psi(x)=0
\]
for all $\psi\in C^\infty (H)$ and all $x\in H$. Then ${\cal
W}\sqsubseteq {\cal Y}$ at $(\bar x, \bar p)$.
\end{prop}
\begin{proof}
Let us consider the set of $s+v$ operators ${\cal U}:=({\cal W}_1,
\ldots, {\cal W}_s, {\cal V}_1, \ldots, {\cal V}_v)$. Since
$S({\cal U},{\cal Y})=0$ in $H$ and
\[
\det \frac\partial {\partial U}({\cal M}_{({\cal U,Y})}S)_C (\bar
U, \bar Y) \neq 0 \ ,
\]
where $\bar U=(\bar W, \bar V)$, we have that ${\cal U}$ is
algebraically dependent on ${\cal Y}$ at $(\bar x, \bar p)$.
Moreover, considering the vector of correlations
\[
T_i(W,U)=W_i-U_i\ ,\quad i=1, \ldots, s \ ,
\]
we have that ${\cal W}$ is algebraically dependent on ${\cal U}$
everywhere. Therefore ${\cal W}\sqsubseteq {\cal Y}$ at $(\bar x,
\bar p)$ according to definition \ref{contain2}.
\end{proof}

We are now going to show that the statement of proposition
\ref{chiave} remains valid if the relation of algebraic dependence
is replaced by the one specified by definition \ref{contain2}.
\begin{prop}\label{opensub}
Let ${\cal W}= ({\cal W}_{1}, \ldots, {\cal W}_{s})$ and ${\cal
Y}=({\cal Y}_1,\ldots,{\cal Y}_{r})$ be two sets of operators of
class ${\cal O}_K$, such that ${\cal W}\sqsubseteq {\cal Y}$ at
$(\bar x, \bar p)\in \dominio$. Then there exists a neighborhood
$O\subset \dominio$ of $(\bar x, \bar p)$ such that:
\renewcommand{\theenumi}{\roman{enumi}}
\begin{enumerate}
\item The main parts $MW$ of the symbols of $\wop$ are functions
of the main parts $MY$ of the symbols of $\yop$ in $O$. More
precisely, there exist a neighborhood $\Omega\subset \rn^r$ of
$\bar Y= MY(\bar x, \bar p)$, and a function $f:\Omega \to \rn^s$,
$f\in C^\infty(\Omega)$, such that $MW(x,p)= f(MY(x,p)) \ \forall
\,(x,p)\in O$. \label{opensubi}

\item ${\cal W}\sqsubseteq {\cal Y}$ at all points of $O$.
\label{opensubii}
\end{enumerate}
\renewcommand{\theenumi}{\arabic{enumi}}
\end{prop}
\begin{proof}
Let ${\cal Z}^{(1)}, \ldots, {\cal Z}^{(q)}$ be the sets of
operators required by definition \ref{contain2}. Define ${\cal
Z}^{(0)}:={\cal W}$, ${\cal Z}^{(q+1)}:= {\cal Y}$, and let
$MZ^{(i)}$ denote the symbols of the main parts of $\zop^{(i)}$,
$i=0, 1, \dots, q+1$. According to proposition \ref{chiave}, there
exist $q+1$ open neighborhoods $O_i$ of $(\bar x,\bar p)$ and
$q+1$ functions $f_i$, with $i=1, \ldots,q+1$, such that
$MZ^{(i)}(x,p)= f_i(MZ^{(i-1)}(x,p))\ $ $\forall\, (x,p)\in O_i$,
and ${\cal Z}^{(i-1)}\sqsubseteq {\cal Z}^{(i)}$ at all points of
$O_i$. It follows that $MW(x,p)= f(MY(x,p))\ $ $\forall\, (x,p)\in
O$, and ${\cal W}\sqsubseteq {\cal Y}$ at all points of $O$, where
$O:= \bigcap_{i=1}^{q+1} O_i$ and $f=f_1\circ f_2 \circ \cdots
\circ f_{i+1}$ is the composition of functions $f_1, \dots,
f_{i+1}$.
\end{proof}

Let us now formulate a definition of ``regular dependence'' for
the operators of the set ${\cal W}= ({\cal W}_1, \ldots, {\cal
W}_s)$ at a point $(\bar x, \bar p)\in K\times \mathbb{R}_p^n$.

\begin{defn}[{\bf regular dependence}] \label{implic}
Let us consider a set ${\cal W}$ of $s$ operators of class ${\cal
O}$. Let us suppose that there exists a set ${\cal Y}=({\cal Y}_1,
\ldots, {\cal Y}_r)$ of operators of class ${\cal O}$, with $r<s$,
such that ${\cal W}\sqsubseteq{\cal Y}$ at the point $(\bar x,
\bar p)\in K\times \mathbb{R}_p^n$. In this case we say that the
operators of ${\cal W}$ are {\it regularly dependent} at $(\bar x,
\bar p)$. If the operators ${\cal W}$ are regularly dependent
almost everywhere in $K\times \mathbb{R}_p^n$, we say that they
are {\it globally dependent} in $K\times \mathbb{R}_p^n$. In this
paper, {\it almost everywhere} means in a subset such that the
closure of its complement has zero measure.
\end{defn}
\begin{defn}
We define the {\it rank} of the set of operators ${\cal W}$ at
$(x,p)$ as the minimum integer $r$ such that there exists another
set ${\cal Y}=({\cal Y}_1, \ldots, {\cal Y}_r)$ with ${\cal
W}\sqsubseteq {\cal Y}$ at $(x,p)$. We denote this rank as
$r=\rank {\cal W}(x,p)$.
\end{defn}

If ${\cal W}=({\cal W}_1, \ldots, {\cal W}_s)$, then obviously
$\rank{\cal W}(x,p) \leq s$ at any point $(x,p)\in K\times
\mathbb{R}_p^n$. The operators ${\cal W}$ are regularly dependent
at $(x,p)$ if and only if $\rank{\cal W}(x,p) < s$. It is also
evident that, if ${\cal W}$ and ${\cal Y}$ are two sets of
operators such that ${\cal W}\sqsubseteq {\cal Y}$ at $(x,p)$,
then $\rank {\cal W}(x,p)\leq \rank {\cal Y}(x,p)$.
The following proposition easily follows from statement
\ref{opensubii} of proposition \ref{opensub}.
\begin{prop}\label{nei}
If $\rank{\cal W}(\bar x,\bar p) =r$, then there exists a
neighborhood $O\subset \dominio$ of $(\bar x, \bar p)$ such that
$\rank{\cal W}(x,p) \leq r \ \forall\, (x,p)\in O$. In particular,
if $\wop$ is regularly dependent at $(\bar x,\bar p)$, then $\wop$
is also regularly dependent at all points of a neighborhood of
$(\bar x,\bar p)$.
\end{prop}

The following proposition is an immediate consequence of
propositions \ref{wxp}.

\begin{prop}
If $s>2n$, then any set ${\cal W} =({\cal W}_1, \ldots, {\cal
W}_s)$ of operators of class ${\cal O}$ is globally dependent.
Equivalently, any $2n+1$ such operators $({\cal W}_1, \ldots,$
${\cal W}_{2n+1})$ are globally dependent.
\end{prop}

The next proposition shows that definition \ref{implic} of regular
dependence indeed includes as a particular case the existence of a
regular correlation among the operators of a set, in the sense of
definition \ref{correl}.
\begin{prop} \label{corind}
If there exists a regular correlation at $(\bar x, \bar p)$ among
the operators ${\cal W} =({\cal W}_1, \ldots, {\cal W}_s)$, then
the operators ${\cal W}$ are regularly dependent at $(\bar x, \bar
p)$.
\end{prop}
\begin{proof}
If the correlation is represented by the noncommutative polynomial
$S\in {\cal S}_N^s$, and $\bar W:= MW(\bar x, \bar p) \in
\mathbb{R}^s$, it follows from condition 1 of definition
\ref{correl} that there exists at least one $j\in \{1, \ldots,
s\}$ such that $(\partial/
\partial W_j)({\cal M}_{\cal W} S)_C(\bar W)\neq 0$. If necessary, let
us rearrange the order of the operators ${\cal W}$ so that
\begin{equation} \label{part}
\frac\partial {\partial W_s}({\cal M}_{\cal W} S)_C(\bar W)\neq 0\
.
\end{equation}
Let us take ${\cal Y}=({\cal Y}_1, \ldots, {\cal Y}_{s-1}) :=
({\cal W}_1, \ldots, {\cal W}_{s-1})$ and consider the set of $s$
correlations
\begin{align*}
S_i(W,Y) &= W_i -Y_i \ ,\quad i=1, \ldots, s-1, \\
S_s(W,Y) &=S(W)\ .
\end{align*}
If $\bar Y:=MY(\bar x, \bar p)$, it is easy to see that
\[
\det\frac{\partial ({\cal M}_{({\cal W},{\cal Y})} S)_C}{\partial
W} (\bar W,\bar Y) = \frac\partial {\partial W_s}({\cal M}_{\cal
W} S)_C(\bar W)\neq 0\ .
\]
Hence ${\cal W}$ is algebraically dependent on ${\cal Y}$ at
$(\bar x, \bar p)$, according to definition \ref{contain}. This
implies that $\wop$ is regularly dependent at $(\bar x, \bar p)$.
\end{proof}

\begin{defn}\label{strongd}
If ${\cal W}= ({\cal W}_{1}, \ldots, {\cal W}_{s})$ is a set of
operators of class ${\cal O}$, let us denote with $r_{\cal
W}(x,p)$ the dimension of the linear space generated by the
differentials $(dMW_1,$ $\ldots, dMW_s)$ of the symbols of the
main parts of the operators ${\cal W}$ at the point $(x,p)\in
K\times\mathbb{R}^n_p$. We have equivalently $r_{\cal W}(x,p)
:=\rank \tilde W(x,p)$, where $\tilde W$ is the $s \times 2n$
Jacobi matrix of the function $MW$. We call $r_{\cal W}(x,p)$ the
{\it main dimension} of the set $\wop$ at the point $(x,p)$.
\end{defn}
Clearly $r_{\cal W}(x,p)\leq s$, and the operators ${\cal W}$ are
quasi-independent at $(x,p)$ if and only if $r_{\cal W}(x,p)=s$
(see definition \ref{strongind}).
\begin{prop}\label{nei2}
If $r_{\cal W}(\bar x,\bar p) =r$, then there exists a
neighborhood $O\subset \dominio$ of $(\bar x, \bar p)$ such that
$r_{\cal W}(x,p) \geq r \ \forall\, (x,p)\in O$. In particular, if
$\wop$ is quasi-independent at $(\bar x,\bar p)$, then $\wop$ is
also quasi-independent at all points of a neighborhood of $(\bar
x,\bar p)$.
\end{prop}
\begin{proof}
The thesis easily follows from the regularity of the symbols of
the operators of class $\oop$.
\end{proof}

The following proposition is an immediate consequence of statement
\ref{opensubi} of proposition \ref{opensub}.
\begin{prop} \label{preind}
Let ${\cal W}= ({\cal W}_{1}, \ldots, {\cal W}_{s})$ and ${\cal
Y}=({\cal Y}_1,\ldots,{\cal Y}_{r})$ be two sets of operators of
class ${\cal O}$, such that ${\cal W}\sqsubseteq{\cal Y}$ at a
point $(\bar x, \bar p)\in \dominio$. Then the vectors $dMW_i(\bar
x, \bar p)$, for $i= 1, \ldots, s$, are linearly dependent on the
vectors $(dMY_1(\bar x, \bar p),$ $\ldots, dMY_r(\bar x, \bar
p))$. Therefore $r_{\cal W}(\bar x, \bar p) \leq r_{\cal Y}(\bar
x, \bar p)$.
\end{prop}

According to statement \ref{opensubii} of proposition
\ref{opensub}, the hypothesis of the preceding proposition
actually implies that ${\cal W}\sqsubseteq{\cal Y}$ at all points
of a neighborhood $O\subset \dominio$ of $(\bar x, \bar p)$. We
have therefore $r_{\cal W}(x, p) \leq r_{\cal Y}(x, p)\ \forall\,
(x,p)\in O$.
\begin{cor}\label{rr}
Let ${\cal W}= ({\cal W}_{1}, \ldots, {\cal W}_{s})$ be a set of
operators of class ${\cal O}$. Then $r_\wop (x,p)\leq \rank
\wop(x, p)\ \forall\, (x,p)\in \dominio$.
\end{cor}
\begin{proof}
If $\rank \wop(x, p)=r$ at $(x, p)\in \dominio$, let ${\cal
Y}=({\cal Y}_1, \ldots, {\cal Y}_r)$ be a set of operators such
that ${\cal W}\sqsubseteq {\cal Y}$ at $(x,p)$. Then proposition
\ref{preind} implies that $r_\wop (x,p)\leq r_\yop (x,p)\leq r$.
\end{proof}

Using proposition \ref{nei} and corollary \ref{rr} one immediately
obtains the theorem which relates the property of regular
dependence with that of quasi-independence of a set of operators.
\begin{thm} \label{indep}
If the operators ${\cal W}$ of class ${\cal O}$ are regularly
dependent at a point $(\bar x, \bar p)\in K \times\mathbb{R}^n_p$,
then there exists a neighborhood $O\subset \dominio$ of $(\bar x,
\bar p)$, such that they are not quasi-independent at any point of
$O$. In particular, they are not globally quasi-independent.
\end{thm}

\begin{cor}
Let us consider the set of $2n$ operators $(x_1, \ldots,x_n, \hat
p_1, \ldots, \hat p_n)$. These operators are quasi-independent.
Therefore they are not regularly dependent at any point $(x,p)$
and are not globally dependent. The same is true for any subset of
this set of operators.
\end{cor}

\begin{prop}
A set ${\cal Y}=({\cal Y}_1, \ldots, {\cal Y}_{2n})$ of operators
of class ${\cal O}_K$ is quasi-independent at a point $(\bar
x,\bar p)\in \dominio$ if and only if ${\cal Y} \sqsupseteq (x_1,
\ldots, x_n, \hat p_1, \ldots, \hat p_n)$ at that point.
\end{prop}
\begin{proof}
Let
\[
{\cal Y}_j= \sum_{|\alpha|\leq m_j}A_{j,\alpha} (x)\hat p^\alpha \
,\quad j=1,\ldots,2n
\]
be the standard representation of the operators ${\cal Y}_j\in
{\cal O}_K$, see definition \ref{standard}. Consider the $2n$
noncommutative polynomials of class $\sop_N^{4n}$
\[
S_j(Y, X, P)= Y_j- \sum_{|\alpha|\leq m_j}A_{j,\alpha}
(X)P^\alpha\ ,\quad j=1,\ldots,2n\ .
\]
We have
\[
(\mop_{(\yop, x, \hat p)}S_j)_C(Y,X,P)= Y_j- MY_j(X,P) \,,
\]
where
\[
MY_j(X,P)= \sum_{|\alpha|= m_j}A_{j,\alpha} (X)P^\alpha
\]
is the symbol of the main part of $\yop_j$ according to definition
\ref{hg}. We see therefore that, if ${\cal Y}$ is a
quasi-independent set at $(\bar x, \bar p)$, then the hypotheses
of definition \ref{contain} are satisfied with ${\cal W} = (x_1,
\ldots, x_n, \hat p_1, \ldots, \hat p_n)$. Hence $(x_1, \ldots,
x_n, \hat p_1, \ldots, \hat p_n)$ is algebraically dependent on
${\cal Y}$ at $(\bar x, \bar p)$.

Viceversa, if ${\cal Y} \sqsupseteq (x_1, \ldots, x_n, \hat p_1,
\ldots, \hat p_n)$ at $(\bar x, \bar p)$, then from proposition
\ref{preind} we have that $r_{\cal Y}(\bar x, \bar p)\geq 2n$, so
that the set ${\cal Y}$ is quasi-independent at $(\bar x, \bar
p)$.
\end{proof}

\begin{rem}
We have considered two main local conditions on a set of
operators: quasi-independence and regular dependence. A
justification for their names is provided by theorem \ref{indep},
which shows that these conditions are mutually incompatible. One
can however still ask if these conditions embrace all possible
cases. Not surprisingly, the answer to this question is negative.
We prove in fact in Appendix \ref{appa} that there exist sets of
operators which are neither quasi-independent nor regularly
dependent. This suggests that it might be possible to find
improvements on the definitions of dependence and independence for
sets of operators, so as to reduce or even suppress the gap
between the two conditions. The achievement of such a goal is left
for future investigations.
\end{rem}

\subsection{Integrable set of operators and integrable operator}
\label{iqs}

In this section we give the definition of integrable operator
$\hop\in {\cal O}$: it is based on the concept of integrable set
of operators of class $\oop$. We give also a non-formal definition
of quantum system and integrable quantum system. These definitions
will be often employed in the following papers of this series.
They are motivated by their correspondence with the analogous
concepts of classical mechanics, which will be now briefly
recalled (see reference \cite{TMMO}).

\begin{defn}
Let $A$ and $B$ be two differentiable functions defined on the
$2n$-dimensional symplectic manifold $\dominio$. Their {\it
Poisson bracket} $\{A, B\}$ is the function defined as
\begin{equation} \label{pbra}
\{A, B\}:= \sum_{i=1}^n \left (\frac {\partial A}{\partial p_i}
\frac{\partial B}{\partial x_i} - \frac {\partial B}{\partial p_i}
\frac{\partial A}{\partial x_i}\right)\ .
\end{equation}
\end{defn}

We recall that a set of functions $V=(V_1,\ldots,V_l)$, $V_i:M \to
\rn$, $i=1, \dots, l$, is said to be {\it functionally
independent} if the differentials $dV_1,\ldots,dV_l$ are linearly
independent almost everywhere in $M$.

\begin{prop} \label{strind2}
Let $(V_1,\ldots,V_l)$ and $(W_1,\ldots,W_s)$ be two sets of
functionally independent functions on a $2n$-dimensional
symplectic manifold, such that $\{V_i,W_k\} =0$ for $i=1,\ldots,l$
and $k=1,\ldots,s$. Then $l+s\leq 2n$.
\end{prop}

\begin{defn}\label{clint}
A set $V=(V_1, \ldots, V_k; V_{k+1}, \ldots, V_{2n-k})$ of
functionally independent functions on a $2n$-dimensional
symplectic manifold $M^{2n}$ is said to be a {\it (classically)
integrable set on $M^{2n}$, with $k$ central functions $V_1,
\ldots, V_k$}, if $\{V_i,V_j\} =0$ everywhere in $M^{2n}$ for
$i=1,\ldots,k$ and $j=1,\ldots,2n-k$. If the hamiltonian $H$ of a
dynamical system is in involution with all functions of the set
$V$, we say that the system is {\it globally integrable with $k$
central functions} and with integrable set of invariants $V$. In
this case it is easy to see that $H$ is locally dependent on the
set of central functions, i.e., $H=f(V_1, \ldots, V_k)$.
\end{defn}

The quantum equivalent of the preceding definition is contained in
the two following ones.
\begin{defn} [\bf quasi-integrable operator]\label{defintset}
Let us consider the set of operators of class ${\cal O}_K$
\begin{equation}
{\cal W}=({\cal W}_1, \ldots, {\cal W}_k; {\cal W}_{k+1}, \ldots,
{\cal W}_{2n-k}) \ , \label{intset}
\end{equation}
with $0<k\leq n$ (some of these operators may be functions only of
$x$). Let these operators satisfy the following two conditions:
\begin{enumerate}
\item The $2n-k$ operators ${\cal W}_{1}, \ldots, {\cal W}_{2n-k}$
are quasi-independent at the point $(x,p) \in \dominio$;

\item $[{\cal W}_{i}, {\cal W}_{j}]=0$ for $i=1,\ldots,k$ and
$j=1,\ldots, 2n-k$.
\end{enumerate}
We then say that the operators ${\cal W}_{1}, \ldots, {\cal
W}_{2n-k}$ are an {\it integrable set of operators at the point
$(x,p)$ with $k$ central operators} ${\cal W}_1, \ldots, {\cal
W}_k$.

If in condition 1, instead of independence at a point $(x,p)$, we
have global quasi-independence in the full phase space $\dominio$,
then we say that this set is a {\it (globally) integrable set of
operators on $\dominio$}.

Let $\hop$ be an operator of class ${\cal O}$ and let us suppose
that the set $({\cal H}, {\cal W}_{1}, \ldots, {\cal W}_{k})$ is
globally dependent, where ${\cal W}_{1}, \ldots, {\cal W}_{k}$ are
central operators of an integrable set of operators $\wop$ at a
point $(x,p)$. For example, ${\cal H}=S({\cal W}_{1}, \ldots,
{\cal W}_{k})$, where $S$ is an arbitrary usual polynomial of $k$
variables, that is $S\in {\cal S}_C^{k}$. We suppose also that
$[{\cal H},{\cal W}_{i}]=0$ for each $i=1,\ldots,2n-k$. In this
case, we will say that ${\cal H}$ is a {\it quasi-integrable
operator with $k$ central operators at the point $(x,p)$}. If the
set ${\cal W}$ is globally independent, then we will say that
${\cal H}$ is a {\it (globally) quasi-integrable operator with $k$
central operators}.
\end{defn}

\begin{defn}[\bf quasi-integrable quantum system] \label{qis}
Let ${\cal H}$ be an operator of class ${\cal O}_K$. Let some
solutions of the equation $({\cal H}-\lambda)\psi =0$ describe
phenomena of the microscopic world. By this we mean that they are
(similar to) {\it wave functions} either of real microscopic
systems, or of useful approximated models of these systems. In
this case we will say that ${\cal H}$ defines a {\it (scalar)
quantum system of general type on the configuration space $K$} and
${\cal H}$ is the {\it hamiltonian operator} of the system. If the
hamiltonian ${\cal H}$ is a quasi-integrable operator with $k$
central operators, we will say that the quantum system is {\it
quasi-integrable with $k$ central operators}.
\end{defn}

The symbol of a commutator can be calculated by means of the
following lemma, which is an immediate consequence of proposition
\ref{prod}.
\begin{lem} \label{commut}
If ${\cal A}=\sum_{|\alpha|\leq m_1}A_\alpha (x)\hat p^\alpha$ and
${\cal B}=\sum_{|\alpha|\leq m_2}B_\alpha (x)\hat p^\alpha$, then
\[
([{\cal A},{\cal B}])^{\rm smb}= \sum_{1\leq |\alpha|\leq m} \frac
1 {\alpha !}\left (\frac {\partial^{|\alpha|} A}{\partial
p^\alpha} \frac{\partial^{|\alpha|} B}{\partial x^\alpha} - \frac
{\partial^{|\alpha|} B}{\partial p^\alpha}
\frac{\partial^{|\alpha|} A}{\partial x^\alpha}\right)\ ,
\]
where $A={\cal A}^{\rm smb}$, $B={\cal B}^{\rm smb}$, $\alpha!
=\alpha_1! \cdots \alpha_n!$, $m:= \max(m_1, m_2)$. We have for
any nonvanishing term of this sum
\[
\ord \frac 1 {\alpha !}\left(\frac {\partial^{|\alpha|}
A}{\partial p^\alpha} \frac{\partial^{|\alpha|} B}{\partial
x^\alpha} - \frac {\partial^{|\alpha|} B}{\partial p^\alpha}
\frac{\partial^{|\alpha|} A}{\partial x^\alpha}\right) \leq m_1 +
m_2 -|\alpha|\ .
\]
\end{lem}

By making use of the preceding lemma one can easily prove the
following one, which establishes an important connection between
the symbol of the commutator of two operators, and the Poisson
bracket between the main parts of the respective symbols.
\begin{lem} \label{14}
Let $\ord {\cal A}=k$, $\ord {\cal B}=l$. Then $\ord [{\cal
A},{\cal B}]\leq g$, where $g:=k+l-1$. Moreover, the homogeneous
parts of order $g$ of the commutator and of the Poisson bracket
are connected by the relation
\begin{equation}\label{14f}
(H_g([{\cal A},{\cal B}]))^{\rm smb} = H_g(\{A, B\}) = H_g(\{MA,
MB\})=\{MA,MB\}\ .
\end{equation}
Here $A={\cal A}^{\rm smb}$, $B={\cal B}^{\rm smb}$, and $H_g$
denotes the homogeneous part of order $g$. In particular, if $\ord
[{\cal A},{\cal B}]= g$, then
\[
(M[{\cal A},{\cal B}])^{\rm smb}= \{MA,MB\}\ .
\]
In these formulas, $M[{\cal A},{\cal B}]$ denotes the main part of
the operator $[{\cal A},{\cal B}]$, whereas $MA$ and $MB$ denote
the main parts of symbols $A$ and $B$ respectively.
\end{lem}

\begin{prop} \label{clint2}
Let ${\cal W}=({\cal W}_1, \ldots, {\cal W}_k; {\cal W}_{k+1},
\ldots, {\cal W}_{2n-k})$ be an integrable set of operators on
$K$, and let $MW=(MW_1, \ldots, MW_k; MW_{k+1}, \ldots,
MW_{2n-k})$ denote the set of the main parts of their symbols,
$W_i = \wop^{\rm smb}_i$ for $i=1, \dots, 2n-k$. Then $MW$ is a
classically integrable set on $\dominio$.
\end{prop}
\begin{proof}
According to definitions \ref{defintset} and \ref{strongind}, the
set $MW=(MW_1, \ldots, MW_k;$ $MW_{k+1}, \ldots, MW_{2n-k})$ is
functionally independent on $M^{2n}=\dominio$. Moreover, applying
formula (\ref{14f}) we obtain that $\{MW_i,MW_j\} =0$ everywhere
in $\dominio$ for $i=1,\ldots,k$, $j=1,\ldots,2n-k$.
\end{proof}

The following proposition represents the quantum analogue of
proposition \ref{strind2}.
\begin{prop} \label{strind}
Let $({\cal V}_1,\ldots,{\cal V}_l)$ and $({\cal W}_1,\ldots,{\cal
W}_s)$ be two sets of quasi-independent operators, such that
$[{\cal V}_i,{\cal W}_k] =0$ for $i=1,\ldots,l$, $k=1,\ldots,s$.
Then $l+s\leq 2n$.
\end{prop}
\begin{proof}
According to definition \ref{strongind} of quasi-independence, the
sets $(MV_1, \dots, MV_l)$ and $(MW_1, \dots, MW_s)$ of the main
parts of the symbols of the given operators are functionally
independent. Moreover, it follows from lemma \ref{14} that
$\{MV_i,MW_k\} =0$ for $i=1,\ldots,l$, $k=1,\ldots,s$. The thesis
then follows from proposition \ref{strind2}.
\end{proof}
\begin{cor}
Let an operator ${\cal W}_0$ commute with all the $2n-k$ operators
of the integrable set ${\cal W}$, see formula (\ref{intset}). Then
the operators $({\cal W}_0,{\cal W}_1,\ldots,{\cal W}_{k})$ are
not quasi-independent. Therefore, in definition \ref{defintset},
$k$ is the maximal number of operators of the set ${\cal W}$ which
commute with all operators of this set ${\cal W}$.
\end{cor}
The last corollary motivates the definition \ref{defintset} of
central operators.

There exists some trick which usually allows one to obtain the
property of quasi-independence for a set of commuting operators.
Let ${\cal A}=\sum_{|\alpha|\leq m}A_\alpha (x)\hat p^\alpha$ be
an operator of class ${\cal O}={\cal O}_n$, i.e., $x=(x_1, \dots,
x_n)$. Let us associate with this operator the operator ${\cal B}=
\Phi_{\cal A}$, ${\cal B} \in {\cal O}_{n+1}$, where ${\cal B} :=
\sum_{|\beta|= m}B_\beta (x)\hat P^\beta$, $x=(x_0, x_1, \dots,
x_n)$, $\hat P := (\hat p_0, \hat p_1, \dots, \hat p_n)$, $\beta=
(\beta_0, \beta_1, \dots, \beta_n)$. Here $B_\beta (x) := A_\alpha
(x)$, $\alpha= \alpha(\beta) := (\alpha_1, \dots, \alpha_n)$, with
$\alpha_i := \beta_i$, $i=1, \dots, n$. This embedding ${\cal A}
\mapsto \Phi_{\cal A}$ maps any operator $\aop$ of class ${\cal
O}_n$ to an operator $\bop = \Phi_{\cal A}$ of class ${\cal O}_{n
+1}$, homogeneous with respect to $\hat p$, whose coefficients are
independent of $x_0$. Note that we have ${\cal B}=M{\cal B}$. It
is easy to check that the following propositions are true.

\begin{prop}
For any operators ${\cal A}, {\cal B} \in {\cal O}_n$, we have
$[\Phi_{\cal A}, \Phi_{\cal B}]= \Phi_{[{\cal A}, {\cal B}]}$ and
$[\Phi_{\cal A}, \hat p_0]= 0$.
\end{prop}

\begin{prop}
Let the set of $k$ functions $(F_1, \dots, F_k)$ be locally
independent, where $F_i$ is the symbol of an operator ${\cal F}_i$
of class ${\cal O}_n$ for $i= 1,\dots, k$. Then the set of $k+1$
operators $(\hat p_0, \Phi_{{\cal F}_1}, \dots, \Phi_{{\cal
F}_k})$ is locally quasi-independent.
\end{prop}

\begin{cor}
Let $({\cal F}_1, \dots, {\cal F}_k; {\cal F}_{k+ 1}, \dots, {\cal
F}_{2n -k})$ be a set of operators of class ${\cal O}_n$ such that
$[{\cal F}_i, {\cal F}_j] =0$ for $i=1, \dots, k$, $j=1, \dots,
2n-k$. Let the set $(F_1, \dots, F_{2n-k})$ of the symbols of
these operators be globally functionally independent, i.e., their
differentials are linearly independent almost everywhere. Then the
set of operators $(\hat p^0, \Phi_1, \dots, \Phi_k; \Phi_{k+1},
\dots, \Phi_{2n-k})$, where $\Phi_i = \Phi_{{\cal F}_i}$ for $i=1,
\dots, 2n-k$, is a quasi-integrable set of operators.
\end{cor}

Note that the eigenvalue equation $({\cal A}- \lambda)u =0$ for
the original operator ${\cal A}$, where $u= u(x_1, \dots, x_n)$,
takes the form $(\Phi_{\cal A} - \lambda \hat p_0^m) U =0$ for the
operator $\Phi_{\cal A}$, where $U= U(x_0, x_1, \dots, x_n)$. To
the eigenfunction $u(x)$ there corresponds the eigenfunction of
special form $U(x_0, x)= e^{x_0}u(x)$. Note also that the symbol
$F= \sum_{|\alpha|\leq m}A_\alpha (x) p^\alpha$ of an operator
${\cal F}$ contains more monomials than its main part $MF=
\sum_{|\alpha|= m}A_\alpha (x) p^\alpha$. Hence we can expect that
the probability of not being quasi-independent is essentially
lower for the set of operators $(\hat p^0, \Phi_1, \dots,
\Phi_{2n-k})$ than for the set $({\cal F}_1, \dots, {\cal F}_{2n
-k})$.

\begin{rem} \label{heat}
The main object of the investigations of the present article is
quantum mechanics. We underline however that the class of
operators which we have been considering in this section includes
arbitrary (scalar) linear differential operators which may also be
used in other domains of mathematical physics. An example is given
by the operator
\[
{\cal H}=\frac \partial {\partial t}-\triangle_\xi +U(\xi)\ ,
\]
which corresponds to the heat equation
\[
\frac \partial {\partial t}\psi =\triangle_\xi \psi -U(\xi)\psi\ ,
\]
where $\xi=(\xi_1,\ldots, \xi_n)$, $\triangle_\xi$ is the Laplace
operator on $\xi$, and $\psi=\psi(t, \xi)$. If we take
$x=(t,\xi)$, then we obtain that the operator ${\cal H}$ of $n+1$
variables $x$ belongs to the class ${\cal O}$ considered above. In
case $U(\xi)= U_1(\xi_1) + \cdots + U_n(\xi_n)$, the operator
${\cal H}$ is an operator with separable variables. Moreover, it
is obvious that ${\cal H}$ is a quasi-integrable operator in the
straight Liouville sense, that is with $k=n+1$ (see definition
\ref{qis}).

Any (scalar) quantum system is apparently described by special
solutions of a linear differential equation with nonconstant real
coefficients. For example, consider the Schroedinger equation
\[
-i\frac \partial {\partial t}\psi =\triangle_\xi \psi -U(\xi)\psi
\ ,
\]
with $i=\sqrt{-1}$. In order to obtain an operator of class ${\cal
O}$ it  must be converted into the heat equation
\[
\frac \partial {\partial \tau}\chi =\triangle_\xi \chi -U(\xi)\chi
\ ,
\]
with $\chi=\chi(\tau,\xi)$. In quantum applications, we are
interested in complex valued solutions of type
$\chi=\chi(\tau,\xi)$ with $\tau=it$, where $t\in\mathbb{R}$,
$\xi\in \mathbb{R}^n$. In this case the functions
\[
\psi(t,\xi)= \chi(it, \xi)
\]
are wave functions of the quantum system which is described by the
given Schroedinger equation. If we are interested in the diffusion
of heat, we consider instead real valued solutions $\chi(\tau,
\xi)$ in the real variables $(\tau,\xi)$ of the heat equation.
Note finally that we do not consider in this paper operators with
small parameters. In particular, in our investigation, we take the
Planck constant $\hbar$ equal to 1 in the operators of quantum
mechanics.
\end{rem}
We have already observed that we do not know of any general
definition of independence for a set of operators, which fully
correspond to the notion of independence for a set of functions on
a symplectic manifold. Nevertheless, for concrete cases connected
with applications, the condition of quasi-independence is usually
helpful to distinguish between integrable and nonintegrable
quantum systems. Indeed, this fact is true for the familiar
examples of quantum systems considered in section \ref{simplest},
as the following proposition shows.

\begin{prop}
The hamiltonian operators ${\hat H}$ of the quantum systems of the
three examples \ref{kepler}--\ref{gravit} are quasi-integrable,
with central operators which are indicated in these examples.
\end{prop}

\begin{rem}
The system with hamiltonian
\[
{\hat H}=\sum_{i=1}^n \omega_i \frac{\hat p_i^2+x_i^2}2
\]
is called {\it $n$-dimensional quantum harmonic oscillator}. This
system is obviously quasi-integrable with $k=n$, because the set
of $n$ operators
\[
\frac{\hat p_1^2 +x_1^2}2,\ldots, \frac{\hat p_n^2 +x_n^2}2
\]
is quasi-independent. Let us consider the case of complete
resonance, which means that the vector of frequencies
$\omega=(\omega_1, \ldots, \omega_n)$ is proportional to a vector
with integer components: $\omega =ch$, where $c\in \mathbb{R}$,
$\omega_1 \omega_2 \cdots \omega_n\neq 0$, $h=(h_1, \ldots, h_n)
\in \mathbb{Z}^n$. In Part IV it will be shown that in this case
the quantum system is (strongly) integrable with $k=1$, like the
corresponding classical system. In other words, the hamiltonian
${\hat H}$ of the system commutes with $2n-1$ quasi-independent
operators. The proof of this fact is not based on the separation
of variables. For the construction of such an integrable system of
operators we will use the operation of ``symmetrization'' of the
corresponding classical functions. This operation plays the main
role in the construction of quantum integrable systems starting
from classical integrable systems (quantization). This
construction is described in Part II.
\end{rem}

\section*{Acknowledgements}

The authors thank D. Bambusi and L. Galgani for useful comments
and advices.

\appendix

\section{Implications of regular dependence} \label{appa}

Theorem \ref{indep} asserts that a set of regularly dependent
operators is not quasi-independent. It has been proved using the
properties of the main parts of a set of operators, which one can
derive from the condition of regular dependence. However, we want
to show in this appendix that regular dependence has also relevant
implications on the properties of the homogeneous parts of lower
order.

Formula (\ref{top}) gives an expression for the symbol of the
homogeneous part of order $\bar d$ of the operator $S({\cal
G},{\cal F})$, where $S$ is a noncommutative polynomial such that
$\deg_{({\cal F},\gop)}S =\bar d$. In order to write an expression
also for the symbol of the homogeneous part of order $\bar d-1$,
it is useful to introduce preliminarily a short notation to
indicate the ``second-main part'' of an operator or a polynomial.
\begin{defn}
Given an operator $\fop$ of class $\oop$, such that $\ord \fop=
m>0$, we call {\it second-main part of $\fop$} the homogeneous
part of order $m-1$ of $\fop$, see definition \ref{hg}. We denote
this part with the symbol $M'\fop$:
\[
M'\fop:= H_{m-1}(\fop)\,.
\]
Correspondingly, we call {\it second-main part of the symbol $F$}
the function
\[
M'F:= H_{m-1}(F)= (M'\fop)^{\rm smb}\,.
\]
If $\ord\fop =0$, we define $M'\fop:=0$ and $M'F:=0$.
\end{defn}
Note that, at variance with the main part, the second-main part
may be zero also for a nonvanishing operator of arbitrary order.
For example, if $n=1$, $\fop=x\hat p^3 + \hat p+ \cos x$, then
$M\fop =x\hat p^3$, $M'\fop =0$.

\begin{defn}
Let $S\in \sop^{l,r}_N$ be a noncommutative polynomial, and
$w=(w_1, \dots, w_r)$ a set of integers. If $\deg_w S=\bar d>0$
(see definition \ref{hom}), we call {\it second-main part of $S$
with weights $w$} the quasi-homogeneous part of degree $\bar d-1$
of $S$. We denote this part with the symbol $\mop'_w S$:
\[
\mop'_w S:= C_{\bar d-1,w}(S)\,.
\]
If $\deg_w S=0$, we define $\mop'_w S:= 0$.

If $\fop=(\fop_1, \dots, \fop_r)$ is a set of operators of class
$\oop$, then we call {\it second-main part of $S$ with respect to
$\fop$} the polynomial $\mop'_\fop S:=\mop'_w S$, where $w_i=\ord
{\cal F}_i\ \forall \, i=1,\ldots,r$.
\end{defn}

\begin{defn}\label{semi}
Let $A$ and $B$ be two differentiable functions defined on the
$2n$-dimensional symplectic manifold $\dominio$. Their {\it
Poisson semi-bracket} $\{A, B\}^+$ is the function defined as
\begin{equation} 
\{A, B\}^+:= \sum_{i=1}^n \frac {\partial A}{\partial p_i}
\frac{\partial B}{\partial x_i} \ .
\end{equation}
\end{defn}

The following proposition can easily be proved using proposition
\ref{prod}.
\begin{prop}\label{mainp2}
Let $\wop=(\fop,\gop)$ and $S\in {\cal S}_N^{l,r}$ be as in the
hypotheses of proposition \ref{mainp}. Let $\tilde S= (\mop_\wop
S)_C$ and $\tilde S'= (\mop'_\wop S)_C$ denote the abelianizations
of the main and second-main parts, respectively, of polynomial $S$
with respect to $\wop$. Then
\begin{align}\label{top2}
&(H_{\bar d-1}(S({\cal G,F})))^{\rm smb}=\tilde S' (G,MF) +\sum_{i
=1}^r \frac{\partial \tilde S}{\partial F_i} (G,MF) M'F_i
\nonumber \\
&+ \sum_{i,j}A_{ij}(G,MF) \{MF_i, MF_j\}^+
+ \sum_{i,h}B_{ih} (G,MF) \{MF_i, MG_h\}^+\ .
\end{align}
In the above formula, $A_{ij}$ and $B_{ih}$ are functions of $l+r$
real variables, which one can easily construct from the expression
of $\mop_\wop S$. For instance, if $\mop_\wop S$ is the
noncommutative monomial represented in formula (\ref{ncmon}), then
\begin{align*}
\sum_{i,j}A_{ij}(G,MF) \{MF_i, MF_j\}^+ &= \sum_{h<k} \frac
{(M_\beta)_C (G,MF)} {MF_{\beta_h}MF_{\beta_k}} \{MF_{\beta_h},
MF_{\beta_k}\}^+ \,, \\
\sum_{i,h}B_{ih}(G,MF) \{MF_i, MG_h\}^+ &= \sum_{h\leq k} \frac
{(M_\beta)_C (G,MF)} {MF_{\beta_h}Z_{k}} \sum_{j=1}^l \frac
{\partial Z_k} {\partial G_j}\{MF_{\beta_h}, G_{j}\}^+ \,,
\end{align*}
where $(M_\beta)_C$ denotes the abelianization of monomial
$M_\beta$. When $\mop_\wop S$ is a generic quasi-homogeneous
noncommutative polynomial, the expressions of $A_{ij}$ and
$B_{ih}$ follow by linearity from the formulas written above.
\end{prop}

We can now strengthen proposition \ref{preind} in the following
way:
\begin{prop} \label{preind2}
Let ${\cal W}= ({\cal W}_{1}, \ldots, {\cal W}_{s})$ and ${\cal
Y}=({\cal Y}_1,\ldots,{\cal Y}_{r})$ be two sets of operators of
class ${\cal O}$, such that ${\cal W}\sqsubseteq{\cal Y}$ at a
point $(\bar x, \bar p)\in \dominio$. Then there exists a set of
real numbers dependent on $(\bar x, \bar p)$, which we call
$a_{ih}$, $b_{ih}$, $c_{ihk}$ and $d_{ijk}$, with $i,j=1,\ldots,
s$ and $h,k=1,\ldots, r$, such that at the point $(\bar x, \bar
p)$ the differentials of the functions $MW_i$, $M'W_i$, $W_{ij}^+
:=\{MW_{i},MW_j\}^+$ can be expressed as linear combinations of
the differentials of the functions $MY_h$, $M'Y_h$, $Y_{hk}^+ :=
\{MY_{h},MY_k\}^+$ according to the relations
\begin{align}
d(MW_i)&= \sum_{h=1}^r a_{ih} d(MY_h) \,,\label{a1}\\
d(M'W_i)&= \sum_{h=1}^r \left(a_{ih} d(M'Y_h) +
b_{ih}d(MY_h)\right) +\sum_{h,k=1}^r c_{ihk}dY_{hk}^+ \,,\label{a2}\\
dW_{ij}^+ &=\sum_{h,k=1}^r a_{ih}a_{jk} dY_{hk}^+ +\sum_{h=1}^r
d_{ijh}d(MY_h) \label{a3}
\end{align}
for $i,j=1,\ldots, s$. Note that the same coefficients $a_{ih}$
appear in all three relations.
\end{prop}
\begin{proof}
It is straightforward to verify that, if the sets $\wop$ and
$\yop$ satisfy relations (\ref{a1})--(\ref{a3}) at $(\bar x,\bar
p)$, and at the same point the sets $\yop$ and $\yop'$ (in this
order) satisfy relations of the same type (in general with other
coefficients $a'$, $b'$, $c'$ and $d'$ in place of $a$, $b$, $c$,
$d$), then still other relations of the same type hold at $(\bar
x,\bar p)$ between the sets $\wop$ and $\yop'$. Taking into
account this fact, it is immediate to see that it is enough to
prove the proposition in the hypothesis that $\wop$ is
algebraically dependent on $\yop$.

Let then $S=(S_1, \dots, S_s)$ be a set of polynomials of class
$\sop_N^{s+r}$ as in definition \ref{contain}, and consider the
two sets of functions $\tilde S= (\mop_{(\wop, \yop)} S)_C$,
$\tilde S'= (\mop'_{(\wop, \yop)} S)_C$. For all $(x,p)\in H\times
\rn^n_p$ we have
\[
0=\big(H_{\bar d_i}(S_i({\cal W,Y}))\big)^{\rm smb}= \tilde S_i
(MW,MY) \quad \forall\,i=1, \dots, s\,,
\]
where $\bar d_i= \deg_{(\wop, \yop)}S_i$. From this, one can
deduce that there exists a neighborhood $O\subset H\times \rn^n_p$
of $(\bar x, \bar p)$ and a vector function $f$, such that
$MW(x,p)= f(MY(x,p)) \ \forall \,(x,p)\in O$, see proposition
\ref{chiave}. We also have
\begin{equation}\label{tildea}
d(MW_i)(x,p)= \sum_{h=1}^r \tilde a_{ih}(MY) d(MY_h)(x,p) \quad
\forall\, (x,p)\in O\,,
\end{equation}
where $\tilde a_{ih}$, $i=1, \dots, s$, $h=1, \dots, r$, are the
elements of the matrix
\begin{equation*}
\tilde A(MY):= -\left(\frac{\partial \tilde S}{\partial
W}\big(f(MY),MY\big)\right)^{-1} \left(\frac{\partial \tilde
S}{\partial Y}\big(f(MY),MY\big)\right)\,.
\end{equation*}
For $(x,p)= (\bar x, \bar p)$ equality (\ref{tildea}) coincides
with (\ref{a1}), with
\begin{equation}\label{aih}
a_{ih}= \tilde a_{ih}(MY(\bar x, \bar p))= -\sum_{j=1}^s U_{ij}
\left(\frac{\partial \tilde S}{\partial Y}(\bar W,\bar
Y)\right)_{jh}\,.
\end{equation}
In the above equality we introduced the matrix
\[
U:=\left(\frac{\partial \tilde S}{\partial W}(\bar W,\bar
Y)\right)^{-1}\,,
\]
where $\bar W:=MW(\bar x, \bar p)$, $\bar Y:=MY(\bar x, \bar p)$.

Using proposition \ref{mainp2} we also have at all points of
$H\times \rn^n_p$
\begin{align}\label{d-1}
0=\ &\big(H_{\bar d_i-1}(S_i({\cal W,Y}))\big)^{\rm smb} \nonumber
\\
=\ &\tilde S'_i (MW,MY) +\sum_{k=1}^s \frac{\partial \tilde
S_i}{\partial
W_k} (MW, MY) M'W_k \nonumber\\
&+\sum_{k=1}^r \frac{\partial \tilde S_i}{\partial Y_k} (MW, MY)
M'Y_k +\sum_{h,k=1}^s A_{i,hk}(MW, MY) \{MW_h, MW_k\}^+ \nonumber\\
&+\sum_{h=1}^s \sum_{k=1}^r \Big[B_{i,hk}(MW, MY) \{MW_h,
MY_k\}^+ \nonumber\\
&+ C_{i,hk}(MW, MY) \{MY_k, MW_h\}^+ \Big]\nonumber \\
&+ \sum_{h,k=1}^r D_{i,hk}(MW, MY) \{MY_h, MY_k\}^+\quad
\forall\,i=1, \dots, s\,,
\end{align}
where $A_{i,hk}$, $B_{i,hk}$, $C_{i,hk}$, and $D_{i,hk}$ are given
functions of $s+r$ variables. From definition \ref{semi} and
formula (\ref{tildea}) it follows that
\begin{align}
W^+_{hk}:= \{MW_h, MW_k\}^+ &= \sum_{h',k'=1}^r\tilde
a_{hh'} \tilde a_{kk'} Y^+_{h'k'} \,, \label{w+}\\
\{MW_h, MY_k\}^+ &= \sum_{h'=1}^r \tilde a_{hh'} Y^+_{h'k} \,, \nonumber\\
\{MY_k, MW_h\}^+ &=\sum_{h'=1}^r \tilde a_{hh'} Y^+_{kh'} \,,
\nonumber
\end{align}
where $Y^+_{hk}:= \{MY_h, MY_k\}^+$. We can therefore rewrite
(\ref{d-1}) as
\begin{align} \label{d-1b}
&\tilde S'_i (MW,MY) +\sum_{k=1}^s \frac{\partial \tilde
S_i}{\partial W_k} (MW, MY) M'W_k \nonumber \\
&+\sum_{k=1}^r \frac{\partial \tilde S_i}{\partial Y_k} (MW, MY)
M'Y_k + \sum_{h,k=1}^r \overline D_{i,hk}(MW, MY) Y^+_{hk} =0 \,,
\end{align}
where $MW=f(MY)$ and
\[
\overline D_{i,hk}= \sum_{h',k'=1}^r\tilde a_{h'h} \tilde a_{k'k}
A_{i,h'k'}+ \sum_{h'=1}^r \big(\tilde a_{h'h} B_{i,h'k}+ \tilde
a_{h'k} C_{i,h'h}\big) +D_{i,hk} \,.
\]
By calculating the differential of (\ref{d-1b}) we get
\begin{align*}
&\sum_{k=1}^s \frac{\partial \tilde S_i}{\partial W_k} (f(MY), MY)
d(M'W_k) +\sum_{k=1}^r \frac{\partial \tilde S_i}{\partial Y_k}
(f(MY), MY) d(M'Y_k) \\
&+ \sum_{h,k=1}^r \overline D_{i,hk}(f(MY), MY) dY^+_{hk}
+\sum_{h=1}^r G_{i,h}(x,p)d(MY_h)= 0 \,,
\end{align*}
where $G_{i,h}(x,p)$ are given functions. For $(x,p)= (\bar x,
\bar p)$ the above equality implies (\ref{a2}), with
\begin{align*}
b_{ih}&=-\sum_{j=1}^s U_{ij}G_{j,h}(\bar x,\bar p) \,, \\
c_{ihk}&=-\sum_{j=1}^s U_{ij}\overline D_{j,hk}(\bar x,\bar p) \,,
\end{align*}
while $a_{ih}$ is still given by (\ref{aih}).

Finally, formula (\ref{a3}) is obtained in a similar way, by
differentiating (\ref{w+}) at $(\bar x,\bar y)$.
\end{proof}

Let $r_{\wop}(x,p)$ denote the main dimension of a set of
operators $\wop=(\wop_1, \ldots, \wop_s)$ at the point $(x, p)\in
\dominio$, see definition \ref{strongd}. According to theorem
\ref{indep}, the inequality $r_{\wop}(x,p)<s$ is a necessary
condition for the regular dependence of $\wop$ at $(x,p)$. Let us
suppose that $r_{\wop}(x,p)=s-1$. This implies that there exists a
nonvanishing vector $\gamma= \gamma(x,p)\in \mathbb{R}^s$,
univocally determined apart from a multiplicative scalar
coefficient, such that
\begin{equation} \label{linear}
\sum_{i=1}^s \gamma_i d(MW_i)=0 \ .
\end{equation}
The following proposition provides an additional necessary
condition for regular dependence, which involves the second-main
parts $(M'W_1, \dots, M'W_s)$ of the symbols of the operators
$\wop$, together with their main parts $(MW_1, \dots, MW_s)$ and
the Poisson semi-brackets $W^+_{ij}:= \{MW_i, MW_j\}^+$.
\begin{prop}\label{nec}
Let $\wop=(\wop_1, \ldots, \wop_s)$ be a set of operators of class
${\cal O}_K$, such that $r_{\wop}(\bar x,\bar p)=s-1$, where
$(\bar x, \bar p)\in \dominio$. If the set $\wop$ is regularly
dependent at $(\bar x,\bar p)$, then there exists a neighborhood
$O\subset \dominio$ of $(\bar x,\bar p)$ such that $\rank\wop(x,
p)= r_{\wop}(x,p) =s-1$ for all $(x,p)\in O$. Furthermore, at all
points of $O$ the cotangent vector
\[
v=\sum_{i=1}^s\gamma_i d(M'W_i) \ ,
\]
where $\gamma= \gamma(x,p)\neq 0$ satisfies relation
(\ref{linear}), belongs to the linear subspace $L\subseteq
T^*\mathbb{R}^{2n}_{xp}$ spanned by the cotangent vectors
$d(MW_i)$ and $dW^+_{ij}$ for $i,j=1,\ldots, s$.
\end{prop}
\begin{proof}
If the set $\wop$ is regularly dependent at $(\bar x, \bar p)$,
then from $r_{\wop}(\bar x,\bar p)=s-1$ and from corollary
\ref{rr} it follows that $\rank\wop(\bar x, \bar p)=s-1$.
Moreover, according to proposition \ref{nei} there exists a
neighborhood $O'\subset \dominio$ of $(\bar x,\bar p)$, such that
$\rank\wop(x, p)\leq s-1\ \forall\, (x,p)\in O'$. On the other
hand, from $r_{\wop}(\bar x,\bar p)=s-1$ and from proposition
\ref{nei2} it follows that there exists another neighborhood
$O''\subset \dominio$ of $(\bar x,\bar p)$, such that $r_\wop(x,
p)\geq s-1\ \forall\, (x,p)\in O''$. Therefore, applying again
corollary \ref{rr} we have that $r_\wop(x, p)=\rank\wop(x, p)=
s-1\ \forall\, (x,p)\in O:=O'\cap O''$.

Let $\yop=(\yop_1, \ldots, \yop_{s-1})$ be a set of operators of
class ${\cal O}_K$, such that $\wop\sqsubseteq\yop$ at the point
$(x, p)\in O$. We can then rewrite formula (\ref{a1}) of
proposition \ref{preind2} as
\begin{equation}\label{a1b}
d(MW_i)= \sum_{h=1}^{s-1} a_{ih} d(MY_h) \,, \qquad i=1, \dots,
s\,.
\end{equation}
We can always assume that the operators of set $\wop$ have been
ordered in such a way that the differentials $\left( d(MW_1),
\dots, d(MW_{s-1})\right)$ are linearly independent. If we
consider the square matrix
\[
\bar A:= (a_{ih}, i,h=1, \dots, s-1) \,,
\]
it then follows from (\ref{a1b}) that $\det \bar A\neq 0$. We can
thus write
\begin{equation}\label{ia1b}
d(MY_i)= \sum_{h=1}^{s-1} \alpha_{ih} d(MW_h) \,, \qquad i=1,
\dots, s-1\,,
\end{equation}
where
\[
\bar A^{-1}= (\alpha_{ih}, i,h=1, \dots, s-1) \,.
\]
Using formula (\ref{a3}), with $r=s-1$, we then obtain
\begin{equation}\label{ia3}
dY_{ij}^+ =\sum_{h,k=1}^{s-1} \alpha_{ih}\alpha_{jk} \left(
dW_{hk}^+ -\sum_{l,m=1}^{s-1} d_{hkl}\alpha_{lm}d(MW_m)\right)
\end{equation}
for $i,j=1, \dots, s-1$. Formulas (\ref{ia1b})--(\ref{ia3}) imply
that $d(MY_i)\in L$ and $dY_{ij}^+ \in L\ \forall\,i,j=1, \dots,
s-1$.

It follows from proposition \ref{preind} that $\yop$ is a
quasi-independent set at $(x,p)$. Hence, equalities (\ref{linear})
and (\ref{a1b}) imply that $\sum_{i=1}^s \gamma_i a_{ih}=0\
\forall\,h=1, \dots, s-1$. If we multiply by $\gamma_i$ both
members of formula (\ref{a2}), with $r=s-1$, and then sum over $i$
from 1 to $s$, we thus obtain
\[
v= \sum_{i=1}^s \gamma_i \left( \sum_{h=1}^{s-1} b_{ih}d(MY_h)
+\sum_{h,k=1}^{s-1} c_{ihk}dY_{hk}^+ \right) \,,
\]
whence $v\in L$.
\end{proof}

It is easy to construct examples of sets of operators
$\wop=(\wop_1, \ldots, \wop_s)$ such that $r_{\wop}(\bar x,\bar
p)=s-1$, which do not satisfy the necessary condition for regular
dependence expressed by proposition \ref{nec}. Consider for
instance the case $n=1$, $s=2$, $\wop= (\wop_1, \wop_2)$, with
\[
\wop_1= \sum_{i=0}^{l} f_i(x)\hat p^i \,, \qquad \wop_2=
\sum_{i=0}^{m} g_i(x)\hat p^i \ ,
\]
where $l,m\in \mathbb{N}$. Suppose that $f_l(x)\equiv g_m(x)
\equiv 1$. Then $MW_1=p^l$, $MW_2=p^m$, so that $d(MW_1)
=lp^{l-1}dp$ and $d(MW_2) =mp^{m-1}dp$. It follows that $r_\wop
(x,p)=1$ for all $(x,p)$ such that $p\neq 0$, and $\gamma_1
d(MW_1) +\gamma_2 d(MW_2)= 0$, where $\gamma_1= mp^m$,
$\gamma_2=-lp^l$. Let us consider the covector
\begin{align}
v=\ &\gamma_1 d(M'W_1) +\gamma_2 d(M'W_2) \nonumber \\
=\ &p^{l+m-1}[mf'_{l-1}(x)-
lg'_{m-1}(x)]dx \nonumber\\
&+p^{l+m-2}[m(l-1)f_{l-1}(x)- l(m-1)g_{m-1}(x)]dp \ . \label{v}
\end{align}
Since $W^+_{ij}\equiv 0$ for $i,j=1,2$, subspace $L\subseteq
T^*\mathbb{R}^{2}_{xp}$ of proposition \ref{nec} is the subspace
of covectors which are multiple of $dp$. Therefore a necessary
condition, in order for the set $\wop$ to be regularly dependent
at $(\bar x,\bar p)$, with $\bar p\neq 0$, is that the coefficient
of $dx$, on the right-hand side of (\ref{v}), vanishes for all $x$
belonging to an open neighborhood of $\bar x$. This is equivalent
to the condition that in this neighborhood $mf'_{l-1}(x)-
lg'_{m-1}(x)=0$, or
\begin{equation}\label{ml}
mf_{l-1}(x)- lg_{m-1}(x)=c \ ,
\end{equation}
where $c$ is a constant. It follows that, if the functions
$f_{l-1}(x)$ and $g_{m-1}(x)$ do not satisfy the above relation,
then the set $\wop$ is neither quasi-independent nor regularly
dependent.

Let us consider the case $l=m=1$, $\wop_1= \hat p+ f(x)$, $\wop_2=
\hat p+ g(x)$. Then necessary condition (\ref{ml}) for regular
dependence becomes $f(x)= g(x)+c$. In this particularly simple
case, this condition is also sufficient. When it is satisfied, we
have in fact that the set $\wop=(\wop_1, \wop_2)$ satisfies the
regular correlation $\wop_1- \wop_2 -c=0$. Hence, according to
proposition \ref{corind}, the set is regularly dependent.


\begin{thebibliography}{99}

\bibitem{Landau} L. D. Landau and E. M. Lifshitz, {\it Mechanics},
Pergamon (London), 1960.

\bibitem{Landau2} L. D. Landau and E. M. Lifshitz, {\it Quantum Mechanics},
Pergamon (London), 1960.

\bibitem{TMMO} N. N. Nekhoroshev, Action-angle variables and their
generalizations, {\it Trans. Moscow Math. Soc.} {\bf 26}, 180--198
(1972).

\bibitem{Arnold} V. I. Arnold, {\it Mathematical methods of classical
mechanics}, Springer-Verlag (New York), 1978.

\bibitem{fomenko} A. T. Fomenko, {\it Differential Geometry and
Topology}, Consultants Bureau (New York), 1987.

\bibitem{fasso} F. Fass\`o, Superintegrable Hamiltonian Systems:
Geometry and Perturbations, {\it Acta Appl. Math.} {\bf 87},
93--121 (2005).

\bibitem{Gordoc} N. N. Nekhoroshev, Generalizations of Gordon's theorem,
{\it Regular and Chaotic Dynamics} {\bf 7}, 239--247 (2002).

\bibitem{Gord} N. N. Nekhoroshev, Types of integrability on a
submanifold and generalizations of Gordon's theorem, {\it Trans.
Moscow Math. Soc.} {\bf 65}, 169--241 (2005).

\bibitem{berezin} F. A. Berezin, General concept of quantization,
{\it Comm. Math. Phys.} {\bf 40}, 153--174 (1975).

\bibitem{cimapi} R. Cirelli, A. Mani\`a and L.Pizzocchero, Quantum mechanics
as an infinite dimensional Hamiltonian system with uncertainty
structure, Parts I and II, {\it J. Math. Phys.} {\bf 31},
2891--2897 and 2898--2903 (1990).

\bibitem{cipi} R. Cirelli and L. Pizzocchero, On the integrability of quantum
mechanics as an infinite-dimensional Hamiltonian system, {\it
Nonlinearity} {\bf 3}, 1057--1080 (1990).

\bibitem{cigama} R. Cirelli, M. Gatti and A. Mani\`a, The pure state space of
quantum mechanics as Hermitian symmetric space, {\it J. Geom.
Phys.} {\bf 45}, 267--284 (2003).

\bibitem{semenov} M. A. Semenov-Tian-Shansky, Quantum and Classical Integrable Systems,
in {\it Integrability of nonlinear systems, Lecture Notes in
Phys.} {\bf 495}, pp. 314--377, Springer (Berlin), 1997.

\bibitem{enciso} A. Enciso and D. Peralta-Salas, Classical and
quantum integrability of Hamiltonians without scattering states,
{\it Theor. Math. Phys.} {\bf 148}, 1086--1099 (2006).

\bibitem{hieta} J. Hietarinta, Classical versus quantum
integrability, {\it J. Math. Phys.} {\bf 25}, 1833--1840 (1984).

\bibitem{cgm} J. Clemente-Gallardo and G. Marmo, Towards a
definition of quantum integrability, arXiv:0808.3819 (2008).

\bibitem{weigert} S. Weigert, The problem of quantum
integrability, {\it Physica D} {\bf 56}, 107--119 (1992).

\bibitem{burch} J. L. Burchnall and T. W. Chaudy, Commutative
ordinary differential operators, {\it Proc. R. Soc. London} {\bf
118}, 557--583 (1928); {\bf 134}, 471--485 (1932).

\bibitem{chaly} O. A. Chalych and A. P. Veselov, Commutative rings
of partial differential operators and Lie algebras, {\it Commun.
Math. Phys.} {\bf 126}, 597--611 (1990).

\bibitem{hieta98} J. Hietarinta, Pure quantum integrability, {\it Phys. Lett.
A} {\bf 246}, 97--104 (1998).

\bibitem{Gra_Wint} S. Gravel and P. Winternitz, Superintegrability
with third-order integrals in quantum and classical mechanics,
{\it J. Math. Phys.} {\bf 43}, 5902--5912 (2002).

\end{thebibliography}
\end{document}